\documentclass[aps,prb, superscriptaddress,twocolumn,longbibliography]{revtex4-1}
\usepackage{amsmath}
\usepackage{amssymb}
\usepackage{graphicx}
\usepackage[usenames,dvipsnames]{xcolor}
\usepackage{tikz}
\usepackage{pgffor}
\usepackage{verbatim}
\usepackage{bbm}
\usepackage{float}
\usepackage{mathrsfs}
\usepackage{url}
\usepackage[colorlinks, breaklinks=true,linkcolor=red, citecolor=blue, linktocpage=true]{hyperref}

\renewcommand{\>}{\rangle}
\newcommand{\be}{\begin{equation} }
\newcommand{\ee}{\end{equation} }
\newcommand{\ba}{\begin{eqnarray} }
\newcommand{\ea}{\end{eqnarray} }

\newcommand{\bpm}{\begin{pmatrix}}
\newcommand{\epm}{\end{pmatrix}}
\newcommand{\bmm}{\begin{matrix}}
\newcommand{\emm}{\end{matrix}}

\newcommand{\bea}{\begin{eqnarray}}
\newcommand{\eea}{\end{eqnarray}}
\renewcommand{\v}[1]{\boldsymbol{#1}}
\newcommand{\beq}{\begin{equation} }
\newcommand{\eeq}{\end{equation} }
\newcommand{\non}{\nonumber }

\begin{document}

\title{Solvable models for neutral modes in fractional quantum Hall edges}

 \author{Chris Heinrich}
\author{Michael Levin}
\affiliation{{Department of Physics, James Franck Institute, University of Chicago, Chicago, IL 60637, USA}}


\begin{abstract}
We describe solvable models that capture how impurity scattering in certain fractional quantum Hall edges can give rise to a neutral mode --- i.e. an edge mode that does not carry electric charge. These models consist of two counter-propagating chiral Luttinger liquids together with a collection of discrete impurity scatterers. Our main result is an exact solution of these models in the limit of infinitely strong impurity scattering. From this solution, we explicitly derive the existence of a neutral mode and we determine all of its microscopic properties including its velocity. We also study the stability of the neutral mode and show that it survives at finite but sufficiently strong scattering. Our results are applicable to a family of Abelian fractional quantum Hall states of which the $\nu = 2/3$ state is the most prominent example.
\end{abstract}


\maketitle

\tableofcontents

\section{Introduction}

One of the most important properties of quantum Hall states is that they have gapless edge modes. Every state has at least one such mode, but the structure of these modes varies from state to state. For example, the Laughlin states are believed to have a single chiral edge mode,\cite{wen-cll} while integer quantum Hall states have multiple chiral edge modes --- one for every filled Landau level.\cite{halperin-iqhedge} 

A particularly interesting edge theory is realized by the $\nu = 2/3$ fractional quantum Hall state. This state is believed to have two \emph{counter-propagating} chiral edge modes --- one which looks like the edge mode of a $\nu = 1$ integer quantum Hall state, and one which looks like the edge mode of a $\nu = 1/3$ Laughlin state, but with opposite chirality.\cite{wen-ED,wen-rev,macdonaldedge,johnsonmacdonald23} This edge theory poses a basic puzzle because it naively predicts charge propagation in both directions along the edge, in disagreement with experiment.\cite{CCT} 

A possible resolution to this problem was put forth by Kane, Fisher and Polchinski.\cite{kfandp} In that work, the authors argued that what is missing from the previous picture is impurity-induced electron scattering between the two edge modes. The authors showed that impurity scattering can drive the edge to a special disorder dominated fixed point where one of the edge modes is electrically \emph{neutral} while the other carries charge; the charge mode propagates in the direction determined by the external magnetic field while the neutral mode propagates in the opposite `upstream' direction. This mode structure can explain why current flow is only observed in one direction on the $2/3$ edge. It is also consistent with experiments on the $\nu = 2/3$ edge which have found evidence for upstream neutral modes,\cite{bid2010observation} though the picture has been complicated by more recent {studies which suggest that the $2/3$ edge may have multiple charge modes, perhaps as a result of edge reconstruction.\cite{sabo2017edge} Also, we should mention that other studies have detected neutral modes in quantum Hall states where they were not expected theoretically.\cite{venkatachalam2012local, inoue2014proliferation}}

The main theoretical justification for the neutral mode proposal of Ref.~\onlinecite{kfandp} comes from a renormalization group (RG) analysis of the fixed point edge theory which shows that the fixed point has no relevant perturbations. This calculation proves that the fixed point has a finite basin of attraction; as long as the edge lies in this basin of attraction, impurity scattering will drive the system to the fixed point with a neutral mode. 

While this analysis is powerful, it leaves some important questions unanswered. In particular, it does not give a microscopic picture for how a neutral mode emerges from impurity scattering. In this paper, we seek to provide such a {picture} in the context of concrete models. 

{The} models we consider are built out of two counter-propagating chiral Luttinger liquids together with a collection of discrete impurity scatterers. Our main result is an exact solution of these models in the limit of infinitely strong impurity scattering, which we obtain using a formalism introduced in Ref.~\onlinecite{quadham}. From this solution, we explicitly derive the existence of a neutral mode and we determine all of its microscopic properties including its velocity. Importantly, we also study the stability of the neutral mode and we show that it survives at finite, but sufficiently strong impurity scattering, {as long as this scattering has a random spatial dependence}.

Our results apply to a particular class of fractional quantum Hall (FQH) edge theories of which the $\nu = 2/3$ edge is a special case. Specifically, the edge theories that we analyze are those described by a K-matrix\cite{wen-rev} of the form $K = \bpm k_1 & 0 \\ 0 & - k_2 \epm$ where $k_1$ is an odd integer and $k_2 = k_1 + 2$.\footnote{Much of our analysis applies to general odd $k_1$, $k_2$ but our most important conclusions rely on the assumption that $|k_2-k_1|=2$, for reasons explained in Appendix \ref{Degeneracy}.} These edge theories correspond to a class of Abelian quantum hall states with filling fraction $\nu = \frac{1}{k_1} - \frac{1}{k_2}$. The $\nu = 2/3$ edge corresponds to the case $k_1 = 1$ and $k_2 = 3$.

The structure of the paper is as follows. In section \ref{sec:mpc} we present the models that we study and we summarize our main results. In section \ref{sec:TMIU} we solve our simplest model --- a minimal toy model --- in the infinite scattering limit and derive the existence of a neutral mode. In section \ref{sec:finiteU} we study the toy model with finite but large impurity scattering and we show that the neutral mode survives in this case. Finally, in section \ref{sec:universality} we consider more general and realistic models and we show that our main results still hold. We conclude in section  \ref{sec:conclusion} and mention a few directions for future work.

\begin{figure}[tb]
\includegraphics[width=.35\textwidth]{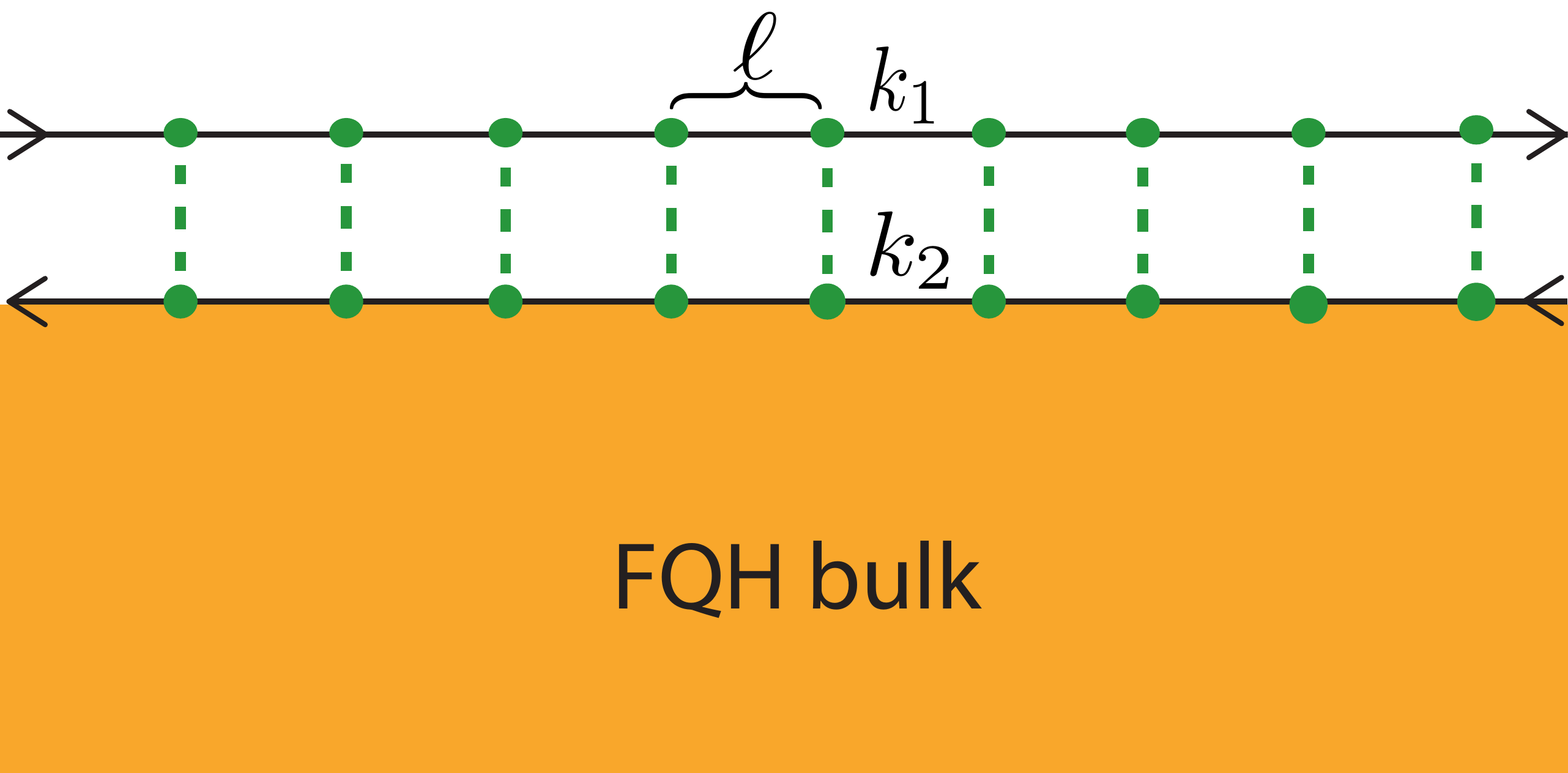}
\centering
\caption{Toy model for a FQH edge with impurity scattering: two counter-propagating chiral Luttinger liquids with parameters $k_1$ and $k_2$ together with a regular lattice of impurity scatterers with spacing $\ell$.}
\label{fig:mpc}
\end{figure}

\section{Models and main results}\label{sec:mpc}

As we mentioned previously, we focus our analysis on FQH edge theories that have filling fraction $\nu = \frac{1}{k_1} - \frac{1}{k_2}$ and that are described by a K-matrix of the form $K = \bpm k_1 & 0 \\ 0 & - k_2 \epm$ where $k_1$ is an odd integer and $k_2 = k_1 + 2$. In the absence of impurity scattering, the edges of these states can be modeled as two counterpropagating chiral Luttinger liquids which look like the edge modes of the $1/k_1$ and $1/k_2$ Laughlin states, but with opposite chiralities.\cite{wen-rev} Our goal is to study how impurity scattering in these systems can produce a neutral mode using concrete models.

We start with a minimal toy model which consists of two counterpropagating chiral Luttinger liquids together with a periodic lattice of impurity scatterers with lattice spacing $\ell$ (Fig. \ref{fig:mpc}). The Hamiltonian is
\begin{align}\label{hamWithPC}
&H =H_0-U\sum_{j} \cos(k_1\phi_1(j \ell) + k_2\phi_2(j \ell) - \alpha_j), \nonumber \\ 
&H_0=\frac{v}{4\pi}\int_{-\infty}^{\infty} dx\Big( k_1(\partial_x\phi_1)^2+k_2(\partial_x\phi_2)^2\Big) 
\end{align}
where $\phi_1, \phi_2$ obey commutation relations
\begin{align}\label{eq:comm}
[\phi_1(x), \partial_y \phi_1(y)]&=-\frac{2\pi i}{k_1}\delta(x-y)\non \\
 [\phi_2(x),\partial_y \phi_2(y)]&=\frac{2\pi i}{k_2}\delta(x-y)\non \\
 [\phi_1(x),\partial_y \phi_2(y)]&=0
\end{align}
Let us explain the different terms in the Hamiltonian. The first term, $H_0$, is a bosonized representation of the two chiral Luttinger liquid edge modes in a normalization convention where the electron creation operators are $\psi_1^{\dagger}=e^{-ik_1\phi_1}$, and $\psi_2^{\dagger}=e^{ik_2\phi_2}$. The second term --- the sum of cosines --- describes a lattice of impurities that scatter electrons from one mode to the other (Fig. \ref{fig:mpc}). The only parameters in the model are $v$, $U$ and $\{ \alpha_j\}$: $v$ is the velocity of the two edge modes, while $U$ and $\{\alpha_j\}$ describe the amplitude and phase of the electron scattering associated with the $j$'th impurity. Notice that we allow the $\alpha_j$ phases to be different for each impurity, but we take the scattering strength $U$ to be constant for simplicity.

What makes the above model useful is that we can study it in a well-controlled fashion and explicitly see that the impurity-induced electron scattering leads to an emergent neutral mode. First, consider the case where $U = 0$ so there is no scattering. In this case, the resulting edge theory has two decoupled modes, $\phi_1, \phi_2$. Both modes are charge-carrying, since the electron density operator is given by $\rho(x) = \frac{1}{2\pi} (\partial_x \phi_1 + \partial_x \phi_2)$. Next suppose we turn on a small $U$. The mode structure remains qualitatively the same as the $U = 0$ case (assuming the $\alpha_j$ phases are chosen randomly) since it is easy to check that the scattering terms are irrelevant perturbations of the $U =0$ edge theory.\footnote{The critical scaling dimension for perturbations with random coefficients in 1D is $3/2$ (see Ref.~\onlinecite{Giamarchi-random}) while the scaling dimension for the scattering term is $\Delta = (k_1 + k_2)/2$ which is always larger than $3/2$.} 

The more interesting case, and our focus in this paper, is when $U$ is \emph{large}. In this case, we show that one of the low energy modes is charged and the other is neutral. We derive this result in two steps. In the first step, we solve the model exactly in the limit $U \rightarrow \infty$ using the formalism of Ref.~\onlinecite{quadham}. The key point is that in this limit, the impurities act as elastic phonon scatterers similarly to a $\delta$-function potential for non-interacting electrons. Consequently, the periodic lattice of impurities produces a phonon band structure just like a periodic potential for electrons. Working out this phonon band structure, we find that there are two low energy phonon {modes, which} are described by the following low energy Hamiltonian:
\begin{align}
\overline{H}_{\text{eff}} =\frac{\bar{v}}{4\pi}\int_{-\infty}^{\infty}dx\ \left(\frac{1}{\nu}(\partial_x\phi_{\rho})^2+\frac{1}{k_2-k_1}(\partial_x\phi_{\sigma})^2 \right), 
\label{Hlowreal}
\end{align}
Here $\bar{v} = \frac{k_2-k_1}{k_1+k_2}v$ is the velocity of the two modes and $\phi_\rho, \phi_\sigma$ are fields obeying commutation relations
\begin{align}
[ \phi_{\rho}(x),\partial_y \phi_{\rho}(y)] &= -2\pi i \nu \delta(x-y) \nonumber \\
[\phi_{\sigma}(x), \partial_y \phi_{\sigma}(y)] &= 2\pi i (k_2-k_1) \delta(x-y) \nonumber \\
[\phi_\rho(x), \partial_y \phi_\sigma(y)] &= 0
\label{rhosigcomm}
\end{align} 
In addition to the Hamiltonian, we also derive an expression for the (coarse-grained) density $\bar{\rho}$:
\begin{equation}
\bar{\rho}(x) = \frac{1}{2\pi} \partial_x \phi_\rho
\label{rhojreal}
\end{equation}
Eqs. (\ref{Hlowreal}-\ref{rhojreal}) tell us the complete low energy mode structure in the limit $U \rightarrow \infty$. Most importantly, they tell us that there are two decoupled low energy modes, $\phi_\rho$ and $\phi_\sigma$, and that $\phi_\rho$ carries charge while $\phi_\sigma$ is \emph{neutral}.

The second step in our derivation is to study what happens when $U$ is large but finite. We analyze this case by adding correction terms to the $U \rightarrow \infty$ low energy theory (\ref{Hlowreal}). We then investigate the effects of these correction terms using a renormalization group (RG) analysis. In the most realistic case where the $\alpha_j$ are chosen randomly, we find that the correction terms have no effect except to renormalize the velocities of the charge and neutral modes. Hence, for random $\alpha_j$, the charge/neutral mode structure persists at large but finite $U$. This is the main result for the first part of the paper.

In the second part of the paper, we generalize the toy model (\ref{hamWithPC}) in two ways. First, we define $H_0$ using an \emph{arbitrary} velocity matrix $V_{ij}$:
\begin{align}
H_0^{\text{gen}}=\frac{1}{4\pi} \int dx\ \sum_{i,j=1}^2 V_{ij} \partial_x \phi_i \partial_x \phi_j
\label{Hzerogen}
\end{align}
Here $V$ can be any real, symmetric, positive definite $2 \times 2$ matrix. Physically, this extension allows the $\phi_1$ and $\phi_2$ modes to have arbitrary velocities and density-density coupling. Our second extension is to make the impurities randomly distributed, rather than regularly spaced. The total Hamiltonian is then
\begin{align}\label{hamWithPCgen}
&H^{\text{gen}} =H_0^{\text{gen}}-U\sum_{j} \cos(k_1\phi_1(x_j) + k_2\phi_2(x_j) - \alpha_j), 
\end{align}
where $x_j$ is the position of the $j$th impurity. Our main result for this part is that $H^{\text{gen}}$ has a charge and a neutral mode at large $U$, just like the toy model $H$. In other words, our derivation generalizes to a more realistic setup with an arbitrary velocity matrix and randomly distributed scatterers.


\section{Toy model: infinite $U$} \label{sec:TMIU}
In this section we solve the toy model (\ref{hamWithPC}) in the limit of infinite scattering strength, $U \rightarrow \infty$. Our main result is that the system has two low energy modes in this limit: a charge mode and a neutral mode. We show that these modes are described by the low energy theory (\ref{Hlowreal}-\ref{rhojreal}).

\subsection{Review of general formalism}\label{sec:rof}
Our solution of the toy model is based on a general formalism for solving quadratic Hamiltonians with large cosine terms, introduced in Ref.~\onlinecite{quadham}. Below we briefly review some of the central results of this formalism before turning to our specific problem.

Consider a general Hamiltonian of the form
\beq
H=H_0-U\sum_{i} \cos(C_i)
\label{genh}
\eeq
defined on some phase space $\{x_1, p_1, x_2, p_2,...\}$.  $H_0$ is a  quadratic function of position and momentum variables $\{x_1, p_1, x_2, p_2,...\}$ and the $C_i$ are linear functions of these variables. The $C_i$'s can be arbitrary except for two restrictions: (1) $\{C_1,C_2,...\}$ are linearly independent, and (2) $[C_i,C_j]$ is an integer multiple of $2\pi i$ for all $i,j$ (so that the cosine terms commute with one another). Ref.~\onlinecite{quadham} showed how to find the low energy spectrum of Hamiltonians of this kind in the limit $U \rightarrow \infty$. 

The basic idea behind the analysis of Ref.~\onlinecite{quadham} is that the cosine terms act as \emph{constraints} in the limit $U \rightarrow \infty$. These constraints force the arguments of the cosine terms to be locked to integer multiples of $2\pi$ at low energies. When this happens, the low energy spectrum of $H$ can be described by an effective Hamiltonian $H_{\text{eff}}$ acting within an effective Hilbert space $\mathcal{H}_{\text{eff}}$. Importantly, the effective Hamiltonian $H_{\text{eff}}$ is \emph{quadratic} and therefore can be diagonalized using elementary methods.

How do we construct the effective Hamiltonian and Hilbert space? The Hilbert space is easy: $\mathcal{H}_{\text{eff}}$ is the subspace of the original Hilbert space consisting of all states $|\psi\>$ satisfying
\beq
\cos(C_i)|\psi\>=|\psi\>, \ \ \ i=1,2,...
\label{hilbeff}
\eeq
As for the Hamiltonian, Ref.~\onlinecite{quadham} described a simple recipe for simultaneously constructing and diagonalizing $H_{\text{eff}}$. The first step is to find all operators $a$ that are linear combinations of the phase space variables $x_1, p_1,...$ and that satisfy the equations
\begin{align}
[a,H_0]&=Ea+\sum_{j}\lambda_j[C_j,H_0] \label{auxeq} \\
[a,C_i]&=0, \ \ \text{for all $i$} \label{aCp}
\end{align}
where $\lambda_j$ and $E$ are arbitary scalars with $E \neq 0$. The above operators $a$ have a simple physical meaning: they describe creation or annihilation operators for the effective Hamiltonian $H_{\text{eff}}$. The scalar $E$ is the energy of the corresponding mode while the scalars $\lambda_j$ can be thought of as Lagrange multipliers associated with the constraints imposed by the cosine terms.

Once the solutions to (\ref{auxeq}-\ref{aCp}) have been identified, the next step is to separate them into two classes: `annihilation operators' with $E>0$ and `creation operators' with $E<0$. If $a_1, a_2,...$ form a complete set of linearly independent annihilation operators, and $a_1^{\dagger},a_2^\dagger,...$ are the corresponding creation operators, then they should be normalized so that
\beq
[a_k,a_{k'}^{\dagger}]=\delta_{kk'}, \ \ \ [a_k,a_{k'}]=[a^{\dagger}_k,a^{\dagger}_{k'}]=0
\eeq
After these steps have been completed, the effective Hamiltonian $H_{\text{eff}}$ can be written down easily: according to Ref.~\onlinecite{quadham}, $H_{\text{eff}}$ is simply given by\footnote{More precisely, Eq. (\ref{heffdiag}) is only guaranteed to hold if we make the additional assumption that the matrix $\mathcal{Z}_{ij} = \frac{1}{2\pi i} [C_i, C_j]$ has a nonvanishing determinant. This property holds for all the systems discussed in this paper.}
\begin{equation}
H_{\text{eff}}=\sum_{k} E_k a^{\dagger}_k a_k 
\label{heffdiag}
\end{equation}

A cautionary note: while it is tempting to conclude that the energy spectrum of $H_{\text{eff}}$ is identical to that of a collection of harmonic oscillators with frequencies $E_k$, this is not quite correct in general. The reason is that, in many cases $H_{\text{eff}}$ has additional degeneracy, i.e. each {occupation number eigenstate} may be $D$-fold degenerate for some $D$. In this paper we will focus on systems where $D=1$ in order to avoid complications associated with this degeneracy. In fact, this is the reason that we restrict to the case $k_2 = k_1 + 2$ (see Appendix \ref{Degeneracy}).


\subsection{Solving the toy model}\label{subsec:solving}
We now use the above formalism to solve the toy model in the $U \rightarrow \infty$ limit. Here, $H_0$ is defined in Eq. (\ref{hamWithPC}), and the $C_j$'s are defined by
\begin{equation}
C_j = k_1 \phi_1(j \ell) + k_2 \phi_2(j \ell) - \alpha_j
\end{equation}
According to the general formalism, we need to find all operators $a$ satisfying the following properties. First, $a$ should be a linear combination of the phase space variables 
$\{\partial_x\phi_1,\partial_x\phi_2\}$:
\beq \label{aop}
a=\int_{-\infty}^{\infty} dx\ (f(x)\partial_x\phi_1+g(x)\partial_x\phi_2)
\eeq
Second, $a$ should obey Eqs. (\ref{auxeq}-\ref{aCp}) for some $\lambda_j, E$ with $E \neq 0$. Finally, given that Eqs. (\ref{auxeq}-\ref{aCp}) have discrete translational symmetry,\footnote{The only part of the toy model that breaks translational symmetry are the $\alpha_j$ phases, and these do not appear in Eqs. \ref{auxeq}-\ref{aCp}.} $f(x)$ and $g(x)$ should obey the Bloch condition
\begin{equation}
f(x+\ell) = e^{-ik\ell} f(x), \quad g(x+\ell) = e^{-ik\ell} g(x)
\label{bloch}
\end{equation}
where $k$ is in the Brillouin zone $[-\pi/\ell, \pi/\ell]$. 

Our task is thus to solve Eqs. (\ref{auxeq}-\ref{aCp}) and (\ref{bloch}). For clarity, we present the result first and then explain the derivation. In short, what we find is that there are an infinite number of solutions to these equations for each value of $k$ in $[-\pi/\ell, \pi/\ell]$. We label these solutions by $a_{n,k}$ and $E_{n,k}$ where $n$ is an integer index, $n = 0, \pm 1, \pm 2,...$. These solutions take the form
\beq
a_{n,k} = \int_{-\infty}^{\infty} \frac{dx}{\sqrt{|E_{n,k}|}} e^{-ikx} (u_{n,k}(x) \partial_x \phi_1 + w_{n,k}(x) \partial_x \phi_2)
\label{ank}
\eeq
where $u_{n,k}$ and $w_{n,k}$ are periodic functions which we derive below. The energies $E_{n,k}$ are given by
\begin{equation}
E_{n,k} = \frac{n \pi v}{\ell} + \frac{(-1)^n v}{\ell} \arcsin \left( \frac{k_2-k_1}{k_2+k_1}\sin k\ell \right)
\label{Enk}
\end{equation}
The solutions come in pairs with $a_{-n,-k} = a_{n,k}^\dagger$ and $E_{-n,-k} = - E_{n,k}$, with the $a_{n,k}$ operators obeying the standard commutation relations
\begin{equation}
[a_{n,k},a_{n',k'}^\dagger]=\delta(k-k')\delta_{n n'}, \quad E_{n,k} > 0
\label{ankcomm}
\end{equation}
With these results in hand, we can immediately write down the effective Hamiltonian $H_{\text{eff}}$ using the general formalism (\ref{heffdiag}):
\beq\label{heff}
H_{\text{eff}}=\sum_n\int_{-\pi/\ell}^{\pi/\ell}dk\ \Theta(E_{n,k}) \ E_{n,k}  a_{n,k}^\dagger a_{n,k}
\eeq
where $\Theta$ denotes the Heaviside step function. 

Equations (\ref{ank}-\ref{heff}) tell us the complete low energy spectrum of the toy model in the limit $U \rightarrow \infty$. To understand the physical interpretation of this spectrum, note that phonons scatter off the impurities elastically in the limit $U \rightarrow \infty$, since in this limit the cosine terms can be modeled as hard constraints on the $\phi_1, \phi_2$ fields. Thus a lattice of impurities gives rise to a band structure for phonons just as a periodic potential gives rise to a band structure for electrons. The above results are consistent with this physical picture: the operators $a_{n,k}^\dagger, a_{n,k}$ (for $E_{n,k} > 0$) can be thought of as creation and annihilation operators for a phonon in band $n$ with crystal momentum $k$. The energy of this phonon mode is given by $E_{n,k}$.  

One thing that these equations do not tell us is the \emph{degeneracy} of the different energy levels of $H_{\text{eff}}$. As we mentioned in the previous section, the phonon occupation numbers $\{a_{n,k}^\dagger a_{n,k}\}$ are not necessarily a complete set of observables; that is, every phonon occupation state may be $D$-fold degenerate for some $D$. We study this issue in Appendix \ref{Degeneracy} using the general formalism of Ref.~\onlinecite{quadham}. We find that for $|k_2 - k_1| = 2$, the toy model has no degeneracy: $D=1$. In contrast, for $|k_2 - k_1| > 2$ we find that the model has an \emph{extensive} degeneracy, i.e. $D$ grows exponentially with the number of impurities. This degeneracy poses many complications, and is the reason that we restrict our analysis to the case $k_2 = k_1+ 2$.

\begin{figure}[tb]
\includegraphics[width=.7\columnwidth]{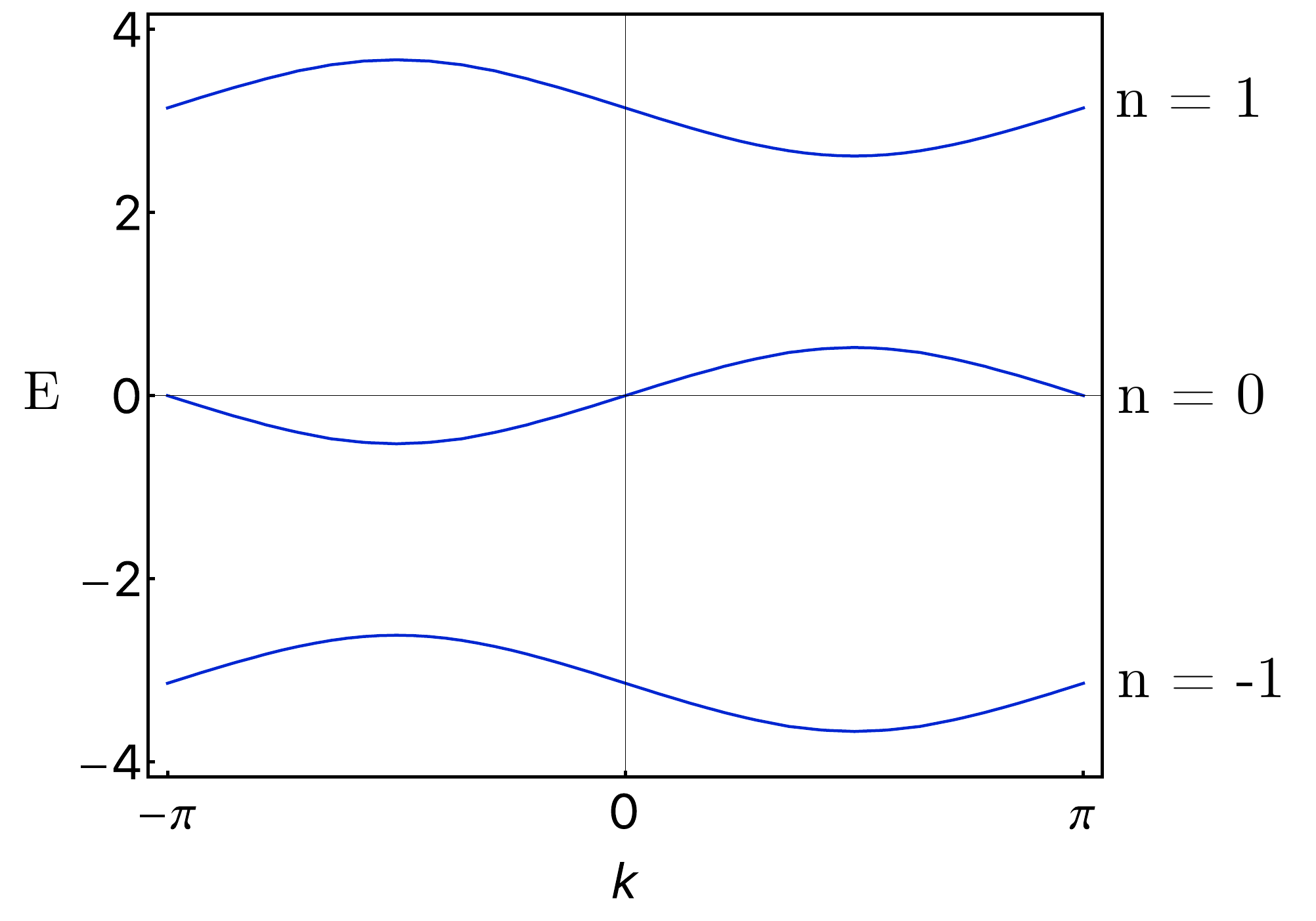}
\centering
\caption{Phonon bandstructure of the toy model in the limit $U \rightarrow \infty$, for the case $k_1 = 1, k_2 = 3$ and $v = \ell = 1$. The {zeros of the $n = 0$ band correspond to low energy phonon modes: $k =0$ corresponds to a right-moving charge mode, while $k = \pi/\ell$ corresponds to a left-moving neutral mode.}}
\label{fig:bandstructure1}
\end{figure}

 
We now solve Eqs. (\ref{auxeq}-\ref{aCp}) and (\ref{bloch}) and derive the results listed above. First, we plug (\ref{aop}) into (\ref{auxeq}), thereby obtaining the differential equations
\begin{align}
f'(x)&=-i\frac{E}{v} f(x)-k_1\sum_{j=-\infty}^{\infty}\lambda_j\delta(x-j \ell) \nonumber  \\
g'(x)&= i\frac{E}{v} g(x)-k_2\sum_{j=-\infty}^{\infty}\lambda_j\delta(x-j \ell)
\label{fgdiff}
\end{align}
Solving this system of equations, we obtain piecewise plane wave solutions of the form
\begin{align}\label{fgdef}
f(x) &= A^{(j)}e^{-i\frac{E}{v}(x-j \ell)}, \nonumber \\ 
g(x) &= B^{(j)}e^{i\frac{E}{v}(x-j \ell)}, \quad \quad j \ell \leq x < (j+1)\ell
\end{align}
To obtain the matching conditions between the $A^{(j)}, B^{(j)}$ coefficients, we note that Eq. (\ref{fgdiff}) implies that
\begin{align}
A^{(j)} &= A^{(j-1)} e^{-i\frac{E \ell}{v}} -\lambda_j k_1 , \nonumber \\
B^{(j)} &= B^{(j-1)} e^{i\frac{E \ell}{v}} -\lambda_j k_2
\end{align}
or equivalently
\beq
\frac{A^{(j)}-A^{(j-1)}e^{-i\frac{E \ell}{v}}}{k_1}=\frac{B^{(j)}-B^{(j-1)}e^{i\frac{E \ell}{v}}}{k_2}
\label{constraint1b}
\eeq 
Another matching condition for $A^{(j)}, B^{(j)}$ comes from the constraint (\ref{aCp}): substituting (\ref{aop}) into (\ref{aCp}), and using an appropriate regularization (see appendix \ref{regularized-delta}), yields
\beq
\frac{A^{(j)}+A^{(j-1)}e^{-i\frac{E \ell}{v}}}{2}=\frac{B^{(j)}+B^{(j-1)}e^{i\frac{E \ell}{v}}}{2}
\label{constraint2b}
\eeq
Using the two constraints (\ref{constraint1b}) and (\ref{constraint2b}), we can solve for $A^{(j)}$ and $B^{(j)}$ in terms of $A^{(j-1)}$ and $B^{(j-1)}$:
\begin{align}
\bpm A^{(j)} \\ B^{(j)} \epm =  T \cdot D(E \ell)  \cdot \bpm A^{(j-1)} \\ B^{(j-1)} \epm
\label{Teq}
\end{align}
where
\begin{equation}
T = \frac{1}{k_2-k_1}\bpm k_2+k_1 & -2k_1 \\
         	          2k_2 & -k_1-k_2 \epm
\label{tmatrixspc}
\end{equation}
and
\begin{equation*}
D(x) = \bpm e^{-ix/v} & 0 \\ 0 & e^{ix/v} \epm
\end{equation*}
Each of the matrices, $T$, $D(E \ell)$ and their product $T \cdot D(E \ell)$, have a simple interpretation. The matrix $T$ can be interpreted as the transfer matrix corresponding to a single impurity: it relates the mode amplitudes just to the right of the impurity to those just to the left. Likewise $D(E \ell)$ can be interpreted as a propagator that describes how the amplitudes change in between the impurities. Finally, $T \cdot D(E \ell)$ can be interpreted as a transfer matrix corresponding to a \emph{unit cell}: it relates the mode amplitudes at the end of the unit cell to those at the beginning of the unit cell.

To proceed further, we impose the Bloch condition (\ref{bloch}), which implies that
\begin{equation}
\bpm A^{(j)} \\ B^{(j)} \epm = e^{-ijk\ell}  \bpm A \\ B \epm
\label{blocheq}
\end{equation}
where $A \equiv A^{(0)}$ and $B \equiv B^{(0)}$. Combining (\ref{Teq}) and (\ref{blocheq}), we arrive at the eigenvalue equation
\begin{equation}
T \cdot D(E \ell) \cdot \bpm A \\ B \epm = e^{-ik\ell}  \bpm A \\ B \epm
\label{eigeq} 
\end{equation}

Equation (\ref{eigeq}) encodes all the information about the phonon band structure and is the main result of our calculation. All that is left is to solve this equation. A quick way to do this is to
note that $\det(T) = -1$ while $\det(D) = 1$, so $\det (T \cdot D) = -1$. It follows that if $T \cdot D$ has an eigenvalue $e^{-ik\ell}$, then its other eigenvalue must be $-e^{ik\ell}$. Hence, $\text{Tr}(T \cdot D)$ must be equal to $-2i \sin(k\ell)$. Comparing this value of the trace with the explicit form of $T \cdot D$, we derive the relation
\begin{equation}
\frac{k_2+k_1}{k_2-k_1}\sin\left(\frac{E\ell}{v}\right)=\sin k\ell
\end{equation}
We can see that for each $k \in [-\pi/\ell, \pi/\ell]$, there are an infinite number of $E$'s that obey this equation. These solutions are precisely the $E_{n,k}$'s given in Eq. \ref{Enk}. The corresponding expressions for $A, B$ can be obtained by straightforward algebra:
\begin{align}
\bpm A_{n,k} \\ B_{n,k} \epm = \bpm k_2 + k_1 + (k_2-k_1) e^{i(k+E_{n,k}/v)\ell} \\ 2k_2 \epm
\end{align}
Putting this all together, we conclude that the most general creation/annihilation operators are of the form (\ref{ank}) where
\begin{align}
u_{n,k}(x) &= \frac{A_{n,k}}{\mathcal{N}_{n,k}} e^{i (k - E_{n,k}/v) \{x\}} \nonumber \\
w_{n,k}(x) &= \frac{B_{n,k}}{\mathcal{N}_{n,k}} e^{i (k + E_{n,k}/v) \{x\}}  
\label{unkwnk}
\end{align}
and where $\{x\}$ is defined to be the distance to the nearest impurity to the left of $x$ (i.e. if $j \ell \leq x < (j+1)\ell$ then $\{x\}= x - j \ell$). The normalization constant $\mathcal{N}_{nk}$ can be determined by demanding that $a_{nk}$ obeys the commutation relations (\ref{ankcomm}):
\begin{equation}
\mathcal{N}_{n,k} = \frac{2\pi}{\sqrt{v}} \left(\frac{|A_{n,k}|^2}{k_1}+\frac{|B_{n,k}|^2}{k_2} \right)^{1/2}
\end{equation}


\subsection{Low energy phonon modes}\label{subsec:low_energy}
The most important feature of the {band structure derived in the previous section} (Fig. \ref{fig:bandstructure1}) is that the {$n=0$} phonon band crosses $E = 0$ in two places: $k=0$ and $k = \pi/\ell$. These crossings imply that the system has two low energy phonon modes with opposite chiralities. {We now derive a low energy Hamiltonian that describes these modes.} 

{In order to be precise, we first need to specify the low energy Hilbert space $\overline{\mathcal{H}}_{\text{eff}}$ for this Hamiltonian. We do this in the obvious way: we} define the Hilbert space $\overline{\mathcal{H}}_{\text{eff}}$ to be the subspace spanned by phonon excitations in the {$n=0$} band with 
\begin{equation*}
|k| \leq \Lambda \quad \text{ or } \quad |k -\pi/\ell| \leq \Lambda
\end{equation*}
where $\Lambda$ is some momentum cutoff with $\Lambda \ll 1/\ell$.

{Likewise, we define the low energy Hamiltonian $\overline{H}_{\text{eff}}$ by projecting $H_{\text{eff}}$ onto the low energy Hilbert space.} The result of this projection is that all the creation and annihilation operators in $H_{\text{eff}}$ drop out except for those with $n=0$ and with $k$ near $0$ or $\pi$. We will relabel these low energy operators as $a_{\rho k}$ and $a_{\sigma k}$ where
\beq
a_{\rho, k} \equiv a_{0,k},\ \ \  a_{\sigma, k} = a_{0, k+\pi/\ell} 
\label{arskdef}
\eeq
and where $|k| \leq \Lambda$. Expressing $H_{\text{eff}}$ in terms of these variables and linearizing the dispersion, we derive the low energy Hamiltonian
\begin{equation}
\overline{H}_{\text{eff}}=\int_{0}^{\Lambda}dk\ \  \bar{v} k(a_{\rho, k}^\dagger a_{\rho, k}+a_{\sigma, -k}^\dagger a_{\sigma, -k})
\label{Hlow}
\end{equation}
where the (renormalized) velocity $\bar{v}$ is given by
\begin{equation}
\bar{v}=\frac{k_2-k_1}{k_1+k_2}v
\label{vrenorm}
\end{equation}
Note that the modes at $k=0$ and $k = \pi/\ell$ have opposite velocities $\pm \bar{v}$.


\subsection{Expression for density operator}\label{subsec:density}

We now derive an expression for the charge density $\rho(x) = \frac{1}{2\pi}(\partial_x\phi_1+\partial_x\phi_2)$ in terms of $a_{\rho, k}$ and $a_{\sigma, k}$. This expression is interesting because it tells us that the $\rho$ ($k = 0$) mode carries \emph{charge} while the $\sigma$ ($k = \pi/\ell$) mode is \emph{neutral}.

The first step is to note that $\rho(x)$ can be expanded as a linear combination of the $a_{n,k}$ operators, that is:
\begin{equation}
\rho(x) =  \sum_n\int_{-\frac{\pi}{\ell}}^{\frac{\pi}{\ell}}dk \ \rho_{n,k}(x) a_{n,k}
\label{rhogeneric}
\end{equation}
Here the $\rho_{n,k}(x)$ are unknown functions that we will determine below. The existence of such an expansion follows from the completeness of the $a_{n,k}$ operators: any linear combination of $\partial_x \phi_1$ and $\partial_x \phi_2$ that commutes with the $C_j$'s can always be expanded in terms of the $a_{n,k}$.\cite{quadham}

Next we find the expansion coefficients $\rho_{n,k}(x)$. To do this, we take the commutator of Eq. (\ref{rhogeneric}) with $a_{-n,-k}$, which gives
\begin{equation}
[\rho(x),a_{-n,-k}] = \rho_{n,k}(x) \cdot \text{sgn}(E_{n,k})
\end{equation}
Evaluating the commutator using the expression for $a_{n,k}$ (\ref{ank}), we obtain
\begin{align}
\rho_{n,k}(x) &= -\frac{i \text{sgn}(E_{n,k})}{\sqrt{|E_{n,k}|}} \non\\
&\times \partial_x \left(e^{ikx} \left[\frac{u_{-n,-k}(x)}{k_1} - \frac{w_{-n,-k}(x)}{k_2} \right] \right) \non \\
&= \frac{\sqrt{|E_{n,k}|}}{v} e^{ikx} \left(\frac{u_{-n,-k}(x)}{k_1} + \frac{w_{-n,-k}(x)}{k_2} \right) 
\end{align} 
where the second equality follows from the differential equation (\ref{fgdiff}). Putting this together, we can write $\rho(x)$ as
\begin{align}
\rho(x) = \sum_n\int_{-\frac{\pi}{\ell}}^{\frac{\pi}{\ell}}dk \ \sqrt{|E_{n,k}|} e^{ikx} z_{n,k}(x) a_{n,k} 
\label{rhojhigh}
\end{align}
where 
\begin{equation}
z_{n,k}(x) = \frac{u_{-n,-k}(x)}{v k_1}  + \frac{w_{-n,-k}(x)}{v k_2}
\end{equation}
At this point, we have found an expression for $\rho(x)$ in terms of the $a_{n,k}$ operators; to complete the calculation we need to go to lower energies and translate Eq. (\ref{rhojhigh}) into an analogous expression for $\rho(x)$ in terms of $a_{\rho, k}$ and $a_{\sigma, k}$. More precisely, since our low energy theory has a momentum cutoff $\Lambda$, we will not be interested in the microscopic density $\rho(x)$, but rather in a \emph{coarse-grained} version of this quantity, which we will denote by $\bar{\rho}(x)$. The coarse-grained density $\bar{\rho}(x)$ is defined by spatially averaging $\rho(x)$ over a region of size $1/\Lambda$.\footnote{The details of this spatial averaging procedure are not important for our purposes: the only property that we will assume below is that $\bar{\rho}$ has identical Fourier components as $\rho$ for wave vectors $|k| \leq \Lambda$ and has vanishing Fourier components for $|k| \gtrsim \Lambda$.} 

Our task is thus to find the expression for $\bar{\rho}(x)$ in terms of $a_{\rho, k}$ and $a_{\sigma, k}$. To this end, we need to \emph{spatially average} the expression for $\rho(x)$ given in Eq. (\ref{rhojhigh}), and then \emph{project} this expression to the low energy Hilbert space $\overline{\mathcal{H}}_{\text{eff}}$. The spatial averaging step can be accomplished by making two changes to Eqs. (\ref{rhojhigh}), namely (1) restricting the integral to $|k| \leq \Lambda$, and (2) replacing $z_{n,k}(x) \rightarrow \bar{z}_{n,k}$, where $\bar{z}_{n,k}$ is defined by averaging $z_{n,k}(x)$ over a unit cell. The projection step can be accomplished by simply throwing out all the terms involving $a_{n,k}$ for $n \neq 0$. 

After performing both steps, the end result is:
\begin{align}
\bar{\rho}(x) &= \int_{-\Lambda}^{\Lambda} dk \ \sqrt{|E_{0,k}|} e^{ikx} \bar{z}_{0,k} a_{0, k} 
\label{rhojint}
\end{align}
The final step is to compute $\bar{z}_{0,k}$. To do this, note that since we are only interested in small $k$ modes, i.e. $|k| \leq \Lambda \ll 1/\ell$, we can make the approximation 
\begin{align}
\bar{z}_{0,k} \approx \bar{z}_{0,0} = \frac{\bar{u}_{0,0}}{v k_1}  + \frac{\bar{w}_{0,0}}{v k_2} 
\end{align}
Similarly, we can approximate $\sqrt{|E_{0,k}|} \approx \sqrt{\bar{v} |k|}$. Subsituting this into Eq. (\ref{rhojint}) and using the expressions for $u_{n,k}, w_{n,k}$ and $\bar{v}$ (\ref{unkwnk}, \ref{vrenorm}), we derive
\begin{align}
\bar{\rho}(x)&=\frac{\sqrt{\nu}}{2\pi}\int_{-\Lambda}^{\Lambda}dk\ \sqrt{|k|} e^{ikx} a_{\rho, k}  
\label{rhoj} 
\end{align} 
Here we have used the identification $a_{0,k}  \equiv a_{\rho, k}$.


\subsection{Charge and neutral modes}
To complete our derivation, we now define two real-space fields $\partial_x\phi_{\rho}$ and $\partial_x\phi_{\sigma}$, which we label the \emph{charge} and \emph{neutral} modes:
\begin{align}
\partial_x\phi_{\rho}(x)&=\sqrt{\nu}\int_{-\Lambda}^{\Lambda}dk\ \sqrt{ |k|} e^{ikx}a_{\rho, k} \\
\partial_x\phi_{\sigma}(x)&=\sqrt{k_2-k_1} \int_{-\Lambda}^{\Lambda}dk \sqrt{ |k|} e^{ikx} a_{\sigma, k}
\label{phirhosigma}
  \end{align}
{One can check that these fields obey the commutation relations}
\begin{align*}
[\phi_{\rho}(x), \partial_y\phi_{\rho}(y)] &= -2\pi i \nu \delta(x-y) \nonumber \\
[\phi_{\sigma}(x), \partial_y \phi_{\sigma}(y)] &= 2\pi i (k_2-k_1) \delta(x-y) \nonumber \\
[\phi_\rho(x), \partial_y \phi_\sigma(y)] &= 0
\end{align*} 
where {the above `$\delta(x)$' is actually} a regularized $\delta$ function that only has Fourier components smaller than $\Lambda$. In terms of these fields, the Hamiltonian (\ref{Hlow}) becomes
\begin{equation*}
\overline{H}_{\text{eff}}=\frac{\bar{v}}{4\pi}\int_{-\infty}^{\infty}dx\ \left(\frac{1}{\nu}(\partial_x\phi_{\rho})^2+\frac{1}{k_2-k_1}(\partial_x\phi_{\sigma})^2 \right),
\end{equation*}
while the (coarse-grained) density operator (\ref{rhoj}) is
\begin{equation*}
\bar{\rho}(x) = \frac{1}{2\pi} \partial_x \phi_\rho
\end{equation*}
{This completes our derivation of the real space low energy theory (\ref{Hlowreal}-\ref{rhojreal}). It also completes our derivation of the neutral mode: indeed, it is obvious that $\phi_\sigma$ is electrically neutral since it does not appear in the above expression for the charge density.}

It is natural to ask: what is the origin of the neutral mode in our calculation? For the above model, this question has a simple answer: the presence of a neutral mode can be traced to {the fact} that the phonon bands cross $E=0$ at both $k=0$ and $k = \pi/\ell$. The key point is that the $k = \pi/\ell$ mode is {\emph{guaranteed}} to be electrically neutral on average, due to its spatial oscillations. 



\section{Toy model: finite $U$}\label{sec:finiteU}

In this section, we analyze the toy model (\ref{hamWithPC}) at large but \emph{finite} scattering strength $U$. Our main result is that the charge/neutral mode structure persists at finite $U$, as long as the $\alpha_j$ phases are chosen {randomly.}


\subsection{RG analysis of low energy theory}

{The key idea behind our analysis is that the low energy effective theory at finite $U$ can be obtained by adding correction terms to the low energy theory at $U = \infty$ (\ref{Hlowreal}). Given this fact, all we have to do is compute these `finite $U$ corrections' and study their effects on (\ref{Hlowreal}). Before doing this, we} first orient ourselves by analyzing the effects of \emph{arbitrary} charge-conserving perturbations on the low energy theory (\ref{Hlowreal}). This will help us distinguish between important and {unimportant corrections}.

We begin by enumerating all local, charge-conserving operators in the low energy theory (\ref{Hlowreal}). To start, it is useful to think about simple examples and `non-examples' of these operators. In particular, we note that the operators $\partial_x \phi_\sigma$ and $\partial_x \phi_\rho$ are valid examples, but $e^{i \text{const.} \cdot \phi_\rho}$ is not since it does not commute with $\int dx \partial_x \phi_\rho$ and therefore breaks charge conservation. Another important example is $e^{i m \phi_\sigma}$. This operator is charge-conserving for all $m$ but it is only a legitimate low energy operator when $m$ is an integer, since it is only in this case that it commutes with the constraints $e^{i C_j}$ that define the low energy Hilbert space (\ref{hilbeff}). One way to see this is to rewrite $\int dx \partial_x  \phi_\sigma$ as
\begin{align}
\int_{-\infty}^{\infty} dx \ \partial_x \phi_\sigma &= \sqrt{k_2-k_1} \int_{-\infty}^{\infty} dx \int_{-\Lambda}^{\Lambda}dk \sqrt{ |k|} e^{ikx} a_{\sigma, k} \nonumber \\
& = \sum_{j} (-1)^j \int_{j \ell}^{(j+1)\ell} dx (k_1 \partial_x \phi_1 + k_2 \partial_x \phi_2) \nonumber \\
&= \sum_{j} 2 (-1)^{j+1} {(C_j+\alpha_j)}
\label{qsigmaid}
\end{align}
(Here the second equality comes from plugging in the definition of $a_{\sigma, k}$ (\ref{ank},\ref{arskdef}) and simplifying). From this identity, we can see that 
\begin{equation}
\Big[\sum_{j} (-1)^{j+1} C_j, \phi_\sigma \Big] = \pi (k_1 - k_2)
\end{equation}
It follows that $\exp(i m\phi_\sigma)$ commutes with $\exp(i\sum_j (-1)^{j+1} C_j)$ only if $m$ is an integer multiple of $2/(k_2 - k_1)$. {Since we specialize} to the case $k_2 - k_1 = 2$, we conclude that $m$ has to be an integer, as claimed above. 

Putting together the above examples, we deduce that the most general charge-conserving operator can be parameterized as 
\begin{equation}
e^{i m \phi_\sigma} f(\{ \partial_x^k \phi_\sigma, \partial_x^l \phi_\rho \})
\label{perturbations}
\end{equation}
where $m$ is an integer and $f$ is a monomial built out of derivatives of $\phi_\sigma$ and $\phi_\rho$. Our next task is to understand the {perturbative} effect of these {operators} on the low energy theory (\ref{Hlowreal}). We do this with a renormalization group (RG) approach. First, we note that the scaling dimension of $e^{i m \phi_\sigma}$ is $\Delta=m^2$ (here we again use the fact that $k_2 -k_1 = 2$). This fact implies that all the {operators} in (\ref{perturbations}) with $|m| \geq 2$ have scaling dimensions larger than $2$ and are thus irrelevant in the RG sense. We can therefore restrict our attention to the {operators} with $m = 0, \pm 1$, of which the only marginal or relevant ones are:
\begin{align}
&\partial_x \phi_\rho, \ \ \partial_x \phi_\sigma, \ \ e^{\pm i\phi_{\sigma}}, \ \ (\partial_x\phi_{\rho})^2, \ \ (\partial_x\phi_{\sigma})^2, \nonumber \\ 
&\partial_x\phi_{\rho}\partial_x\phi_{\sigma},  \ \ e^{\pm i \phi_\sigma} \partial_x \phi_\rho  
\end{align}
Let us consider each of these {perturbations}. The first three terms are unimportant since they can be `gauged away' --- that is, eliminated from the Hamiltonian by an appropriate redefinition of fields. 
This is obvious for $\partial_x \phi_\rho$ and $\partial_x \phi_\sigma$: these terms can be eliminated by completing the square in the Hamiltonian (\ref{Hlowreal}). As for $e^{\pm i \phi_\sigma}$, the fact that this term can be gauged away follows from an observation of Ref.~\onlinecite{kfandp}, namely that when $|k_1 -k_2| = 2$, the three operators $\{\int dx \cos(\phi_\sigma), \int dx \sin(\phi_\sigma), \int dx\partial_x \phi_\sigma\}$ generate an $SU(2)$ symmetry group that leaves the Hamiltonian (\ref{Hlowreal}) invariant. Like any $SU(2)$ generators, these three operators transform like a three component vector under the symmetry that they generate. In particular, this means that we can rotate the operator $\cos(\phi_\sigma)$ into $\partial_x \phi_\sigma$ using the $SU(2)$ symmetry. The latter term can be gauged away, hence $\cos(\phi_\sigma)$ can also be gauged away.

The next two {perturbations}, $(\partial_x\phi_{\rho})^2, (\partial_x\phi_{\sigma})^2$, are also relatively unimportant since their only effect is to shift the charge and neutral mode velocities. Thus, the only {perturbations} we need to worry about {are} $\partial_x\phi_{\rho}\partial_x\phi_{\sigma}$ and $e^{\pm i \phi_\sigma} \partial_x \phi_\rho$. These {perturbations} \emph{do} have an important effect: they couple the charge and neutral modes so that both of the resulting hybridized modes are charge-carrying. \footnote{This hybridization effect is obvious for $\partial_x \phi_\rho \partial_x \phi_\sigma$; to see why it occurs for $e^{\pm i \phi_\sigma}$, note that these three operators form a multiplet under the $SU(2)$ symmetry mentioned above.} Thus, these {perturbations} are dangerous from our perspective because they destroy the {decoupled} charge/neutral mode structure if they are present.

\subsection{Fate of neutral mode}
The next step is to compute the finite $U$ corrections for the impurity model (\ref{hamWithPC}) and determine whether the two `dangerous' perturbations discussed above, namely $\partial_x\phi_{\rho}\partial_x\phi_{\sigma}$ and $e^{\pm i \phi_\sigma} \partial_x \phi_\rho$, are generated. This calculation is technical so we postpone it to the next section, and skip to the main result: what we find is that these perturbations \emph{do} appear as finite $U$ corrections but with spatially dependent coefficients. In particular, $\partial_x\phi_{\rho}\partial_x\phi_{\sigma}$ appears in the form
\begin{equation}
\sum_j (-1)^j \partial_x \phi_\rho  \partial_x \phi_\sigma(j \ell)
\label{rs}
\end{equation}
with a coefficient that {changes} sign every unit cell. Meanwhile $e^{\pm i \phi_\sigma} \partial_x \phi_\rho$ appears in the form
\begin{equation}
\sum_j \cos(\phi_\sigma(j \ell) - \beta_j) \partial_x \phi_\rho(j \ell)
\label{esr}
\end{equation}
where the $\beta_j$ are determined by the original $\alpha_j$ phases via the relation
\begin{equation}
\beta_{j+1} - \beta_j = (-1)^j (\alpha_{j+1} - \alpha_j)
\label{ba}
\end{equation}

The alternating coefficient in Eq. (\ref{rs}) has a very important consequence: it suppresses the effect of $\partial_x\phi_{\rho}\partial_x\phi_{\sigma}$, effectively rendering it \emph{irrelevant}. 
Likewise, the $\beta_j$ phases in (\ref{esr}) can also lead to cancellations that suppress this perturbation, but these cancellations are more delicate and depend on the values of $\alpha_j$. Thus to determine the fate of the charge/neutral mode structure, we need to fix a choice of $\alpha_j$. Here we focus on two possibilities: (a) {$\alpha_j  = j \Phi$ for some $\Phi$}, and (b) random $\alpha_j$. Physically, case (a) corresponds to a situation where an identical amount of magnetic flux $\Phi$ threads between each pair of impurities, between the $\phi_1, \phi_2$ edge modes. Likewise, case (b) corresponds to random magnetic flux and can be thought of as capturing some aspects of a more realistic random impurity system. 

Interestingly these two cases lead to different physics. In the uniform flux case (a), we obtain $\beta_{j+1} - \beta_j = (-1)^j \Phi$, so we can take $\beta_{2j} = 0$ and $\beta_{2j+1} = \Phi$. Substituting this into (\ref{esr}), we see that in the long distance limit, the finite $U$ corrections generate a term of the form 
\begin{equation}
[\cos(\phi_\sigma) + \cos(\phi_\sigma - \Phi)] \partial_x \phi_\rho
\end{equation}
Evidently there is no cancellation (for generic $\Phi$) so the $e^{\pm i \phi_\sigma} \partial_x \phi_\rho$ perturbation is not suppressed. Therefore, the charge and neutral modes will become hybridized at finite $U$. In other words, the charge/neutral mode structure does \emph{not} persist at finite $U$ in this case. 

On the other hand, in the random flux case (b), the $\beta_j$ phases are also random and independent, so the operators $e^{\pm i \phi_\sigma} \partial_x \phi_\rho$ appear with random phases. These random phases make $e^{\pm i \phi_\sigma} \partial_x \phi_\rho$ irrelevant, since {it has a scaling dimension, $\Delta = 2$,} which is larger than the critical dimension of $3/2$ for perturbations with random coefficients.\cite{Giamarchi-random} Therefore, in this case, both of the dangerous {perturbations} are suppressed and hence the charge and neutral mode survive at finite $U$ in this case.

\subsection{Finite $U$ corrections}
To complete the discussion, we need to compute the finite $U$ corrections and derive Eqs. (\ref{rs}) and (\ref{esr}). Before doing this, we first review the general formalism for these {corrections.} 

{In Ref.~\onlinecite{quadham} it was argued that the low energy spectrum of (\ref{genh}) for large, finite $U$} can be obtained by adding appropriate correction terms to the $U = \infty$ effective Hamiltonian $H_{\text{eff}}$ (\ref{heffdiag}). These correction terms can always be written in the following general form:\footnote{This expression holds assuming the matrix $\mathcal{Z}_{ij} = \frac{1}{2\pi i} [C_i, C_j]$ has a nonvanishing determinant, as is the case for all the systems discussed in this paper.}
\begin{equation}
\sum_{\v{m}} e^{i \sum_j m_j \Pi_j} \epsilon_{\v{m}}(\{a_k, a_k^\dagger\})
\label{finUgen})
\end{equation}
Here the sum runs over integer vectors $\v{m} = (m_1, m_2, ...)$ and the $\epsilon_{\v{m}}$ are some unknown functions of $\{a_{k},a_{k}^\dagger\}$ which also depend on $U$. Also, $\Pi_j$ is defined by
\begin{equation}
\Pi_j= \frac{1}{2\pi i} \sum_i \mathcal{N}^{-1}_{ji} [C_i,H_0]
\end{equation}
where $\mathcal{N}$ is the matrix $\mathcal{N}_{ji} = -\frac{1}{(2\pi)^2} [C_j, [C_i, H_0]]$. Note that (\ref{finUgen}) does not tell us the functional form of $\epsilon_{\v{m}}(a_k, a_k^\dagger)$: this is system dependent and cannot be determined without more calculation.

To understand where the expression (\ref{finUgen}) comes from, note that when $U$ is finite, we expect that there is a small amplitude for the system to tunnel between the minima of the cosine terms, i.e. $C_j\rightarrow C_j-2\pi m_j$. Thus, the corrections to $H_{\text{eff}}$ should be a sum of the most general possible operators describing tunneling processes of this kind. Eq. (\ref{finUhi}) is precisely such a sum of (general) tunneling operators. Indeed, one can see that (\ref{finUgen}) gives a matrix element for the tunneling process $C_j\rightarrow C_j-2\pi m_j$ using the commutation relation 
\begin{equation}
[C_j, \Pi_i] = 2\pi i \delta_{ji}
\end{equation}
(See Ref.~\onlinecite{quadham} for more details).

We now apply the above general formalism to the lattice impurity model (\ref{hamWithPC}). For simplicity, we start with the case where only one of the impurities has a finite value of $U$ while the others have $U = \infty$. In this case, we only have to think about the finite $U$ corrections associated with a single impurity --- say, the $j$th impurity. Thus, the general expression (\ref{finUgen}) reduces to:
\begin{equation}
\sum_{m = -\infty}^{\infty} e^{i m \Pi_j} \epsilon_{m}(\{a_{n,k}, a_{n,k}^\dagger\})
\label{finUhi}
\end{equation}
where $\Pi_j$ is defined by
\beq
\Pi_j=2\pi i \frac{[C_j,H_0]}{[C_j,[C_j,H_0]]}
\label{Pip}
\eeq
and where $\epsilon_{m}$ are some unknown functions of $\{a_{n,k},a_{n,k}^\dagger\}$ which also depend on $U$. \footnote{Here the reason that the $\Pi_j$ operator takes a simpler form is that the matrix $\mathcal{N}_{ji} = -\frac{1}{(2\pi)^2} [C_j, [C_i, H_0]]$ is diagonal.} 

Equivalently, the finite $U$ corrections can be written in the real space form
\begin{equation}
\sum_{m = -\infty}^{\infty} e^{i m \Pi_j} f_{m}(\partial_x^k \phi_1(j \ell), \partial_x^l \phi_2(j \ell)) 
\end{equation}
where the function $f_m$ obtained by expressing $\epsilon_{m}(\{a_{n,k}, a_{n,k}^\dagger\})$ in terms of $\partial_x \phi_1, \partial_x \phi_2$.

Next, consider the case where \emph{all} the impurities have the same finite value of $U$. For large $U$, we expect the dominant corrections to be independent tunneling processes associated with single impurities. Therefore, in this limit, we expect the finite $U$ corrections to be a sum of the single impurity corrections (\ref{finUhi}) over all $j$:
\begin{equation}
\sum_{m, j} e^{i m \Pi_j} f_{m}(\partial_x^k \phi_1(j \ell), \partial_x^l \phi_2(j \ell)) 
\label{finUhi2}
\end{equation}

{Our main task is to translate the correction terms (\ref{finUhi2}) into the low energy theory with two linearly dispersing phonon modes (\ref{Hlowreal}). We} start with the operator $f_{m}(\partial_x^k \phi_1, \partial_x^l \phi_2)$. To translate this operator into the low energy theory, we note that $\partial_x \phi_1$ and $\partial_x \phi_2$ are linearly related to $a_{n,k}, a_{n,k}^\dagger$, which are in turn linearly related to $\partial_x \phi_\rho$ and $\partial_x \phi_\sigma$. Hence the $f_{m}$ operator corresponds to some function of the derivatives of $\phi_\rho$ and $\phi_\sigma$, evaluated at $j \ell$. Next, consider the operator $e^{i\Pi_j}$. Translating this operator into the low energy theory requires more sophisticated arguments. First, we use the relation $[C_i, \Pi_j] = 2\pi i \delta_{ij}$ together with the identity (\ref{qsigmaid}) to deduce that 
\begin{equation}
\left[\int dx \partial_x \phi_\sigma, e^{i\Pi_j} \right] = 4\pi (-1)^j e^{i \Pi_j}
\label{commepi}
\end{equation}
Writing down the most general charge-conserving operator in the low energy theory that is consistent with these commutation relations, we derive
\begin{equation}
e^{i \Pi_j} = e^{i (-1)^j (\phi_\sigma(j \ell) - \beta_j)} (1 + ...)
\label{epi1}
\end{equation}
where $\beta_j$ is some unknown phase and the `$...$' includes terms built out of derivatives of $\phi_\sigma, \phi_\rho$. To fix the value of the $\beta_j$ phases, or more precisely, the \emph{relative} values of these phases, consider the operator 
\begin{equation}
\mathcal{O} = \Pi_j + \Pi_{j+1} + C_{j} - C_{j+1} + \alpha_j - \alpha_{j+1} 
\end{equation}
The operator $\mathcal{O}$ has two important properties: (i) it is linear in the fields $\partial_x \phi_1, \partial_x \phi_2$, and (ii) it commutes with $C_j$ for all $j$. (Here the second property follows from the commutation relation {$[C_j, C_i] = i\pi (k_2-k_1) \cdot \text{sgn}(j-i)$)}. Given these two properties, it follows that $\mathcal{O}$ can be expanded as a linear combination of $a_{n,k}, a_{n,k}^\dagger$ since $a_{n,k}, a_{n,k}^\dagger$ form a complete basis for the set of operators satisfying (i), (ii).\cite{quadham} This means that we have
\begin{equation}
\Pi_j + \Pi_{j+1} = \alpha_{j+1} - \alpha_j + C_{j+1} - C_j + \sum_{n,k} ( \lambda_{n,k} \cdot a_{n,k} + h.c)
\end{equation}
for some constants $\lambda_{n,k}$. If we now exponentiate both sides of this equation and take the ground state expectation value in the limit $U \rightarrow \infty$, we see that 
\begin{equation}
\text{arg}(\<e^{i \Pi_j} e^{i \Pi_{j+1}}\>) = \alpha_{j+1} - \alpha_j 
\end{equation}
since $e^{ i C_j} = e^{i C_{j+1}} = 1$ in this limit. Comparing this result to the expression (\ref{epi1}), we deduce that $\beta_{j+1} - \beta_{j} = (-1)^j (\alpha_{j+1} - \alpha_j)$ 
as in Eq. (\ref{ba}). 

Putting this all together, we conclude that the finite $U$ corrections (\ref{finUhi2}) take the following form in the low energy theory: 
\begin{equation}
\sum_{m , j} e^{i m (-1)^j (\phi_\sigma(j \ell) - \beta_j)} \tilde{f}_{m}(\partial_x^k \phi_\rho(j \ell), (-1)^j \partial_x^l \phi_\sigma(j \ell))
\label{finUlow}
\end{equation}
Here $\beta_j$ is given by Eq. (\ref{ba}) and $\tilde{f}_m$ are some unknown functions. The reason for the factor of $(-1)^j$ multiplying $\partial_x^l \phi_\sigma$ is that $\phi_\sigma$ describes a mode near $k = \pi/\ell$, and therefore the relation between $\phi_\sigma$ and $\phi_1, \phi_2$ alternates sign at every impurity. The same reasoning explains why there is no factor of $(-1)^j$ multiplying $\partial_x^k \phi_\rho$ since $\phi_\rho$ describes a mode near $k = 0$.

From Eq. (\ref{finUlow}), we can immediately read off the correction terms that are proportional to the two `dangerous' perturbations, $\partial_x \phi_\rho \partial_x \phi_\sigma$ and $e^{ \pm i \phi_\sigma} \partial_x \phi_\rho$. Specifically, we can see that $\partial_x \phi_\rho \partial_x \phi_\sigma$ appears in the $m = 0$ terms, and takes the form $\sum_j (-1)^j \partial_x \phi_\rho (j \ell) \partial_x \phi_\sigma(j \ell)$. Likewise, $e^{ \pm i \phi_\sigma} \partial_x \phi_\rho$ appears in the $m = \pm 1$ terms and takes the form $\sum_j \cos(\phi_\sigma(j \ell) - \beta_j) \partial_x \phi_\rho(j \ell)$. This completes our derivation of Eqs. (\ref{rs}) and (\ref{esr}).


\section{Generalized models}\label{sec:universality}

Thus far we have focused on the toy model (\ref{hamWithPC}). This model has several special (and unrealistic) properties: (i) the impurities are arranged in a perfect lattice{, and} (ii) the two modes $\phi_1$ and $\phi_2$ move at the same speed $v$ and are decoupled from one another. We now investigate whether the charge and neutral modes persist under more realistic conditions. We build up to the most realistic case in several steps. First, in section \ref{genlattice}, we consider what happens when the velocity matrix is arbitrary, the impurities form a lattice with an arbitrary unit cell, and $U \rightarrow \infty$. Then, in section \ref{rand}, we consider the case where the velocity matrix is arbitary, the impurities are \emph{randomly} positioned, and $U \rightarrow \infty$. Finally, in section \ref{subsubsec:randfinU}, we consider the most realistic case of an arbitrary velocity matrix, random impurities and a finite $U$.

\subsection{General impurity lattices}\label{genlattice}
\begin{figure}[tb]
\includegraphics[width=.35\textwidth]{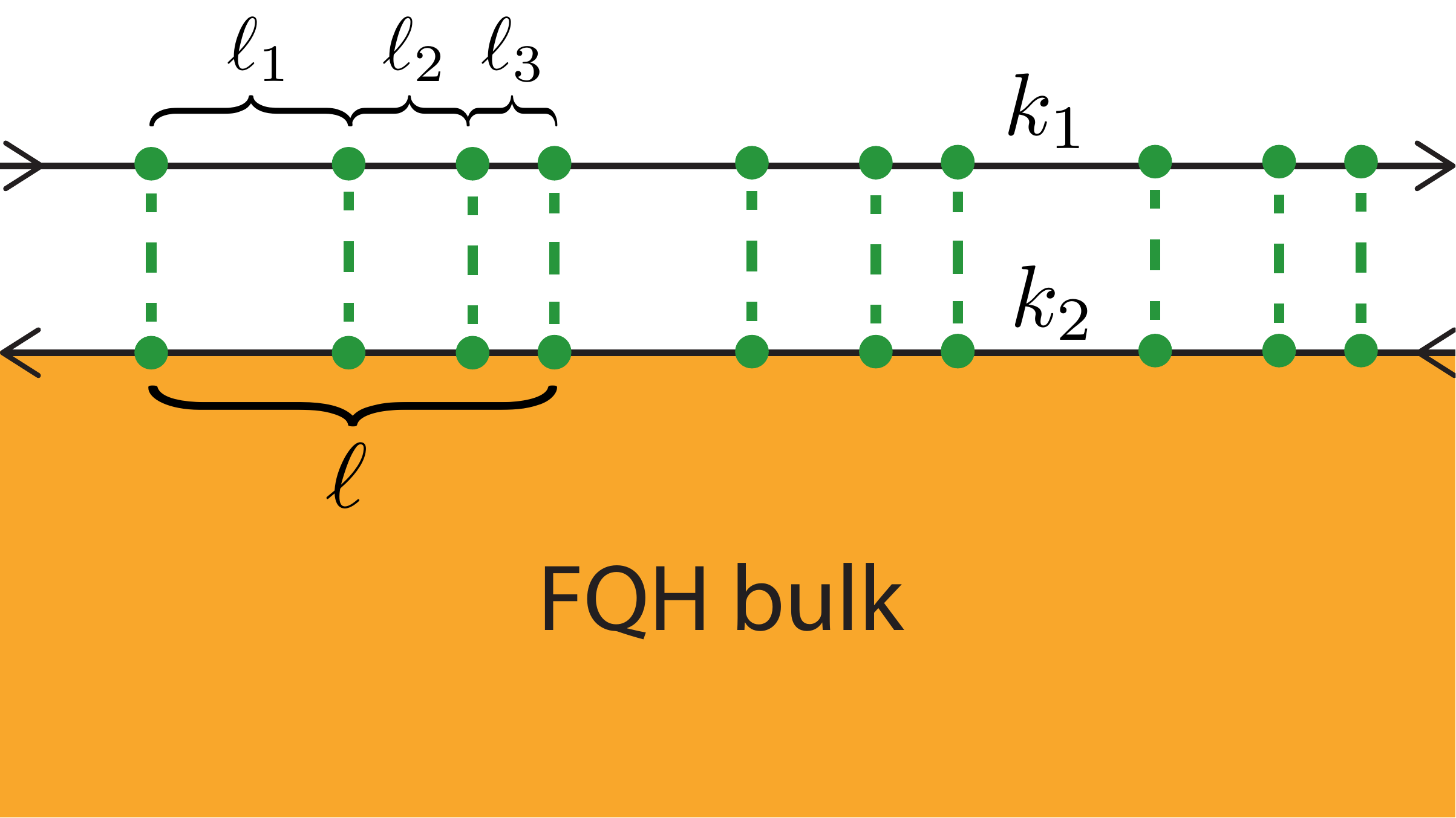}
\centering
\caption{ Generalized impurity lattice model with three impurities per unit cell. The unit cell has length $\ell$ while the spacings between the impurities are $\ell_1, \ell_2$ and $\ell_3$. }
\label{general_lattice}
\end{figure}

In this section we generalize the toy model in two ways. First, instead of focusing on the simplest possible impurity lattice, with only one impurity per unit cell, we consider a \emph{general} lattice with $m$ impurities in a unit cell of length $\ell$ with arbitrary spacing $\ell_1,...,\ell_m$ (Fig. \ref{general_lattice}). Second, instead of assuming that the two modes $\phi_1$ and $\phi_2$ are decoupled from one another and move with the same speed $v$, we consider an arbitrary velocity matrix $V_{ij}$. That is, we consider a Hamiltonian of the form $H^{\text{gen}}$ (\ref{hamWithPCgen}), with the impurities arranged in a general lattice. 

\subsubsection{Structure of low energy modes}
We begin by analyzing the phonon modes for these more general systems. Our main result is that when $U \rightarrow \infty$ these systems have two low energy phonon modes, whose creation/annihilation operators we denote by $a_{\rho, k}$ and $a_{\sigma, k}$ (see below for their definitions). These modes are described by an effective Hamiltonian of the form
\begin{equation}
\overline{H}_{\text{eff}}=\int_{0}^{\Lambda}dk\ \  (v_\rho k a_{\rho, k}^\dagger a_{\rho, k}+ v_\sigma k a_{\sigma, -k}^\dagger a_{\sigma, -k})
\label{hlowgen}
\end{equation}
where $v_\rho, v_\sigma > 0$ are defined below and $\Lambda \ll 1/\ell$ is a momentum cutoff.

The calculation is very similar to the one for the toy model. Indeed, Bloch's theorem guarantees that the phonon creation and annihilation operators $a_{n,k}$ take the same form as before:
\begin{equation}
a_{n,k} = \int \frac{dx}{\sqrt{|E_{n,k}|}} e^{-ikx} (u_{n,k}(x) \partial_x \phi_1 + w_{n,k}(x) \partial_x \phi_2)
\label{ank2}
\end{equation}
where $k$ takes values in the Brillouin zone $[-\pi/\ell, \pi/\ell]$ and $u_{n,k}(x), w_{n,k}(x)$ are periodic functions with period $\ell$. The effective Hamiltonian also takes the same form as before:
\begin{equation*}
H_{\text{eff}}=\sum_n\int_{-\pi/\ell}^{\pi/\ell}dk\ \Theta(E_{n,k}) \ E_{n,k}  a_{n,k}^\dagger a_{n,k}
\end{equation*}
Thus, all we have to do is find the phonon energies $E_{n,k}$ and the Bloch functions $u_{n,k}(x), w_{n,k}(x)$. Proceeding in exactly the same way as in section \ref{subsec:solving}, these quantities can be obtained by solving an eigenvalue equation of the form given in Eq. (\ref{eigeq}):
\begin{equation}
T_{\text{cell}}(E) \cdot \bpm A \\ B \epm = e^{-ik\ell}  \bpm A \\ B \epm
\label{eigeqgencell}
\end{equation}
where $T_{cell}$ is the transfer matrix associated with a single unit cell. The only difference from the toy model is that the transfer matrix $T_{\text{cell}}$ is more complicated due to the fact that the unit cell contains $m$ impurities, and the velocity matrix is more general. In particular, $T_{\text{cell}}$ is given by
\begin{equation}
T_{\text{cell}}(E) = T D(E \ell_m) T \cdots T D(E \ell_1)
\label{Tcellgen}
\end{equation}
where $D(x) = e^{-i W x}$ and $W = K V^{-1}$ and  $K = \bpm k_1 & 0 \\ 0 & -k_2 \epm$.

\begin{figure}[tb]
\includegraphics[width=.5\textwidth]{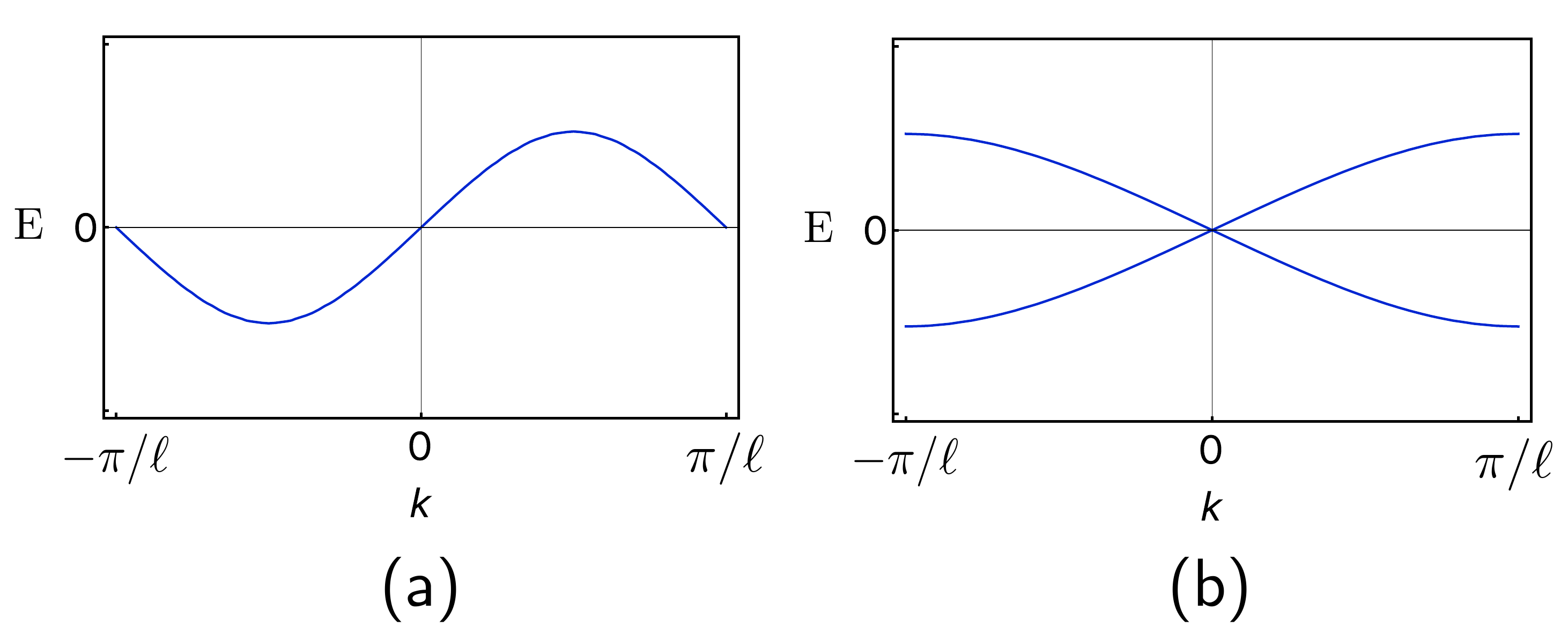}
\centering
\caption{A schematic figure illustrating the difference between band structures with an odd and even number of {impurities} per unit cell. {In the odd case (a), the phonon bands have zeros at $k=0$ and $k=\pi/\ell$. In the even case (b), the phonon bands have both zeros at $k=0$.}}
\label{bs_odd_even}
\end{figure}

Eq. (\ref{eigeqgencell}) tells us the entire phonon band structure, but for our purposes, we only need to understand the \emph{low energy} phonon modes. Therefore, in what follows we will focus on solving (\ref{eigeqgencell}) in the limit of small $E$. To this end, we expand $T_{\text{cell}}(E)$ to linear order in $E$. Using the fact that {$T^2 = \mathbbm{1}$}, we obtain:
\begin{equation}
T_{\text{cell}}(E) = \begin{cases} 
T - i E (T W \ell_{\text{odd}} + W T \ell_{\text{even}}) & \ \text{if $m$ is odd} \\
\mathbbm{1} - i E (W \ell_{\text{odd}} + T W T \ell_{\text{even}}) & \ \text{if $m$ is even} 
\end{cases}
\label{Tcellevenodd}
\end{equation}
where
\begin{align}
\ell_{\text{odd}} &=\ell_1+\ell_3+\cdots  \nonumber \\ 
\ell_{\text{even}} &=\ell_2+\ell_4+\cdots 
\label{levenodd}
\end{align}
From these expressions, we can readily compute the eigenvectors and eigenvalues of $T_{\text{cell}}$. First suppose $m$ is odd. In this case, perturbation theory gives the following 
eigenvalues for $T_{\text{cell}}$:
\begin{equation}
1 - i E \ell/v_{\rho}, \quad -1 - i E \ell/v_{\sigma}
\end{equation}
where
\begin{align}
v_\rho  = \frac{2}{\text{Tr}(TW + W)}, \quad v_\sigma = \frac{2}{\text{Tr}(TW - W)}
\label{vrhosigma}
\end{align}
Substituting these expressions into Eq. (\ref{eigeqgencell}), we see that there are two low energy phonon modes, which are located near $k=0$ and $k = \pi/\ell$ and have velocities $v_\rho$ and $-v_\sigma$ respectively (Fig. \ref{bs_odd_even}(a)). 

Similarly, when $m$ is even, perturbation theory gives the following eigenvalues for $T_{\text{cell}}$:
\begin{equation}
1 - i E \ell/v_{\rho}, \quad 1 + i E \ell/v_{\sigma}
\end{equation}
where $v_\rho^{-1}$ and $-v_{\sigma}^{-1}$ are the two eigenvalues of the matrix
\begin{equation}
\widetilde{W} = W \ell_{\text{odd}} + T W T \ell_{\text{even}},
\label{Mdef}
\end{equation}
Plugging these expressions into Eq. (\ref{eigeqgencell}), we see that there are {again two} low energy phonon modes, but now both are located near $k=0$ with velocities $v_\rho$ and $-v_\sigma$ (Fig. \ref{bs_odd_even}(b)). 

Combining these results, we see that for either parity of $m$, the lowest energy modes are described by the effective Hamiltonian (\ref{hlowgen}) --- where $a_{\rho, k}$ and $a_{\sigma, k}$ are the creation/annihilation operators for the two low energy modes. Note that the definitions of $a_{\rho, k}$ and $a_{\sigma,k}$ are different depending on whether $m$ is odd or even due to the fact that the modes are located in different places in $k$ space. If $m$ is odd, then 
\begin{equation*}
a_{\rho, k} \equiv a_{0,k}, \quad a_{\sigma,k} \equiv a_{0,k+\pi/\ell}
\end{equation*}
as in Eq. \ref{arskdef}, while if $m$ is even, 
\begin{equation*}
a_{\rho,k} \equiv a_{0,k}, \quad a_{\sigma,k} \equiv a_{1,k}
\end{equation*}
where `$0$' and `$1$' are the band indices for the two bands that pass through $k=0$ and $E = 0$.

\subsubsection{Expression for density operator}
In order to understand how much charge is carried by these low energy modes, we now express the (coarse-grained) density operator $\bar{\rho}(x)$ in terms of $a_{\rho, k}$ and $a_{\sigma, k}$. As in section \ref{subsec:density}, the first step is to express the microscopic density operator $\rho(x) = \frac{1}{2\pi} (\partial_x \phi_1 + \partial_x \phi_2)$ in terms of $a_{n, k}$. This step closely parallels the derivation of Eq. (\ref{rhojhigh}), and the result takes a similar form:
\begin{align}
\rho(x) = \sum_n\int_{-\frac{\pi}{\ell}}^{\frac{\pi}{\ell}}dk \ \sqrt{|E_{n,k}|} e^{ikx} z_{n,k}(x) a_{n,k}
\label{rhohighgen}
\end{align}
where
\begin{equation}
z_{n, k}(x) = \bpm 1 & 1 \epm \cdot V^{-1} \cdot \bpm  u_{-n, -k}(x) \\ w_{-n, -k}(x) \epm
\end{equation}

{As before,} the quantity that we want to compute is the \emph{coarse-grained} density $\bar{\rho}(x)$, obtained by spatially averaging $\rho(x)$ over a length scale of order $1/\Lambda$, where $\Lambda$ is a momentum cutoff much smaller than $1/\ell$. To perform this spatial averaging step, we restrict the integral in (\ref{rhohighgen}) to $|k| \leq \Lambda$, and replace $z_{n,k}(x) \rightarrow \bar{z}_{n,k}$ where $\bar{z}_{n,k}$ is defined by averaging $z_{n,k}(x)$ over a unit cell. This gives:
\begin{align}
\bar{\rho}(x) = \sum_n\int_{-\Lambda}^{\Lambda}dk \ \sqrt{|E_{n,k}|} e^{ikx} \bar{z}_{n,k} a_{n,k}
\end{align}

To complete the calculation, we need to project the above expression to the Hilbert space generated by the low energy phonon modes. This projection step gives a different result depending on whether $m$ is odd or even. If $m$ is odd, then just as in section \ref{subsec:density}, there is only one low energy mode with $|k| \leq \Lambda$, namely $a_{\rho, k}$ ($\equiv a_{0,k}$), so we obtain
\begin{equation}
\bar{\rho}(x) = \int_{-\Lambda}^{\Lambda} dk \ \sqrt{|v_{\rho} k|} e^{ikx}   \bar{z}_{\rho,k} a_{\rho, k} 
\label{rholowodd}
\end{equation}
On the other hand, if $m$ is even, then there are two low energy modes with $|k| \leq \Lambda$, namely $a_{\rho, k}$ ($\equiv a_{0,k}$) and $a_{\sigma, k}$ ($\equiv a_{1,k}$) so we derive
\begin{equation}
\bar{\rho}(x) = \int_{-\Lambda}^{\Lambda} dk \ e^{ikx} (\sqrt{|v_{\rho} k|}  
\bar{z}_{\rho,k} a_{\rho, k} +  \sqrt{|v_{\sigma} k|}  \bar{z}_{\sigma,k} a_{\sigma, k})
\label{rholoweven}
\end{equation}
Here $z_{\rho, k}, z_{\sigma, k}$ are defined by
\begin{align}
z_{\rho, k}(x) &= \bpm 1 & 1 \epm \cdot V^{-1} \cdot \bpm  u_{\rho, -k}(x) \\ w_{\rho, -k}(x) \epm \nonumber \\
z_{\sigma, k}(x) &= \bpm 1 & 1 \epm \cdot V^{-1} \cdot \bpm  u_{\sigma, -k}(x) \\ w_{\sigma, -k}(x) \epm 
\label{zdef}
\end{align}
while $\bar{z}_{\rho,k}$ and  $\bar{z}_{\sigma,k}$ are defined by averaging $z_{\rho,k}(x)$ and $z_{\sigma,k}(x)$ over a unit cell. 

\subsubsection{Conditions for neutral mode} \label{neutral_cond}
With this preparation we are ready to tackle the main question: determining the conditions under which the $\sigma$ mode is electrically neutral. Our main result is that the $\sigma$ mode is neutral in two cases: (a) $m$ is odd, or (b) $m$ is even and 
\beq
 \ell_{\text{odd}}=\ell_{\text{even}}
\eeq
where $\ell_{\text{odd}}$ and $\ell_{\text{even}}$ are defined as in Eq. (\ref{levenodd}).

We start with case (a). This case is quite simple since when $m$ is odd, $a_{\sigma, k}$ does not appear at all in the expression for $\bar{\rho}$ as we can see from Eq. (\ref{rholowodd}). It thus follows immediately that the $\sigma$ mode is neutral in this case. 

Case (b) is more subtle. Indeed, when $m$ is even, $a_{\sigma, k}$ \emph{does} appear in $\bar{\rho}$ (\ref{rholoweven}) so to determine the amount of charge carried by the $\sigma$ mode, we need to compute the coefficient $\bar{z}_{\sigma, k}$ that multiplies $a_{\sigma, k}$. In fact, since we are interested in low energy properties, the relevant quantity is the $k \rightarrow 0$ limit of this coefficient, $\bar{z}_{\sigma, 0}$.

We compute this quantity in three steps. First, we find the eigenvectors of $T_{\text{cell}}(E)$ (\ref{Tcellgen}) in the $E \rightarrow 0$ limit. To this end, recall from Eq. (\ref{Tcellevenodd}) that $T_{\text{cell}}(E)$ can be approximated by
\begin{equation}
T_{\text{cell}}(E) \approx \mathbbm{1} - i E (W \ell_{\text{odd}} + T W T \ell_{\text{even}})
\end{equation}
Conveniently, this expression is easy to diagonalize when $\ell_{\text{odd}} = \ell_{\text{even}}$. Indeed, in this case, one can check that 
\begin{equation}
[T, W \ell_{\text{odd}} + T W T \ell_{\text{even}}] = 0
\label{vancomm}
\end{equation}
since $T^2 = \mathbbm{1}$. It follows that the eigenvectors of $T_{\text{cell}}(E)$ are the same as $T$, namely: $(1 \ 1)^T$ and $(k_1 \ k_2)^T$.

Next, we substitute the above eigenvectors into the {expressions for the Bloch functions, $u, w$. These expressions, which can be derived in a similar fashion to Eqs. (\ref{unkwnk}), are as follows:
\begin{align}
\bpm u(x) \\ w(x) \epm = e^{ik\{x\}} \cdot e^{-i E W (x-x_j)} \cdot \bpm A^{(j)} \\ B^{(j)} \epm, 
\label{uwgen}
\end{align}
where 
\begin{align}
\bpm A^{(j)} \\ B^{(j)} \epm = T D(E \ell_j) \cdots T D(E \ell_1) \cdot \bpm A \\ B \epm 
\label{uwgen2}
\end{align}
Here we assume that $x$ is located between the $j$th and $j+1$st impurities, i.e. $x_j \leq x < x_{j+1}$, and $\{x\}$ is defined by $\{x\}= x - p \ell$, for $p \ell \leq x < (p+1)\ell$.} 

We start with the second eigenvector. Letting $(A \ B) = (k_1 \ k_2)$ and $E = k =0$ in Eqs. (\ref{uwgen}-\ref{uwgen2}) {gives
\begin{equation}
\bpm u(x) \\ w(x) \epm = \pm \bpm k_1 \\ k_2 \epm 
\end{equation}
with the sign \emph{alternating} across each impurity. This alternating sign is due to the fact that $(k_1 \ k_2)^T$ is an eigenvector of $T$ with eigenvalue $-1$. Likewise, letting $(A \ B) = (1 \ 1)$ gives
\begin{equation}
\bpm u(x) \\ w(x) \epm = \bpm 1 \\ 1 \epm 
\end{equation}
Note that the sign does not alternate in this case since $(1 \ 1)^T$ is an eigenvector of $T$ with eigenvalue $+1$.}

{To complete the calculation, we identify $(k_1 \ k_2)$ with the $\sigma$ mode and $(1 \ 1)$ with the $\rho$ mode and then we average} the above Bloch functions over a unit cell and plug them into (\ref{zdef}) to obtain $\bar{z}_{\sigma, 0}$ and $\bar{z}_{\rho, 0}$. We start with $\sigma$: in this case, the averaging step gives $\bar{u}_{\sigma, 0} = \bar{w}_{\sigma, 0} = 0$ since there is \emph{perfect cancellation} between the `$+$' and `$-$' signs due to the fact that $\ell_{even} = \ell_{odd}$. Hence, when we plug this into (\ref{zdef}), we obtain $\bar{z}_{\sigma, 0} = 0$. We conclude that the $\sigma$ mode is neutral in the low energy, long wavelength limit, to lowest order in $k$. For comparison, if we repeat this calculation for the $\rho$ mode, the averaging step gives $\bar{u}_{\rho, k} = \bar{w}_{\rho, k} \neq 0$ since the sign does not alternate in this case. It follows that $\bar{z}_{\rho, 0} \neq 0$, so the $\rho$ mode carries charge in the low energy, long wavelength limit.

\subsection{Random impurities}
In this section, we consider systems with \emph{randomly} distributed impurities. We start with the $U \rightarrow \infty$ case and then consider the case where $U$ is large but \emph{finite}.

\subsubsection{{Infinite $U$}}\label{rand}

Given the results from the previous section, one might expect random impurity systems to have a neutral mode in the limit $U \rightarrow \infty$ since the `even' and `odd' spacings are equal on average. Here we show that this intuition is correct. 

\begin{figure}[tb]
\includegraphics[width=.32\textwidth]{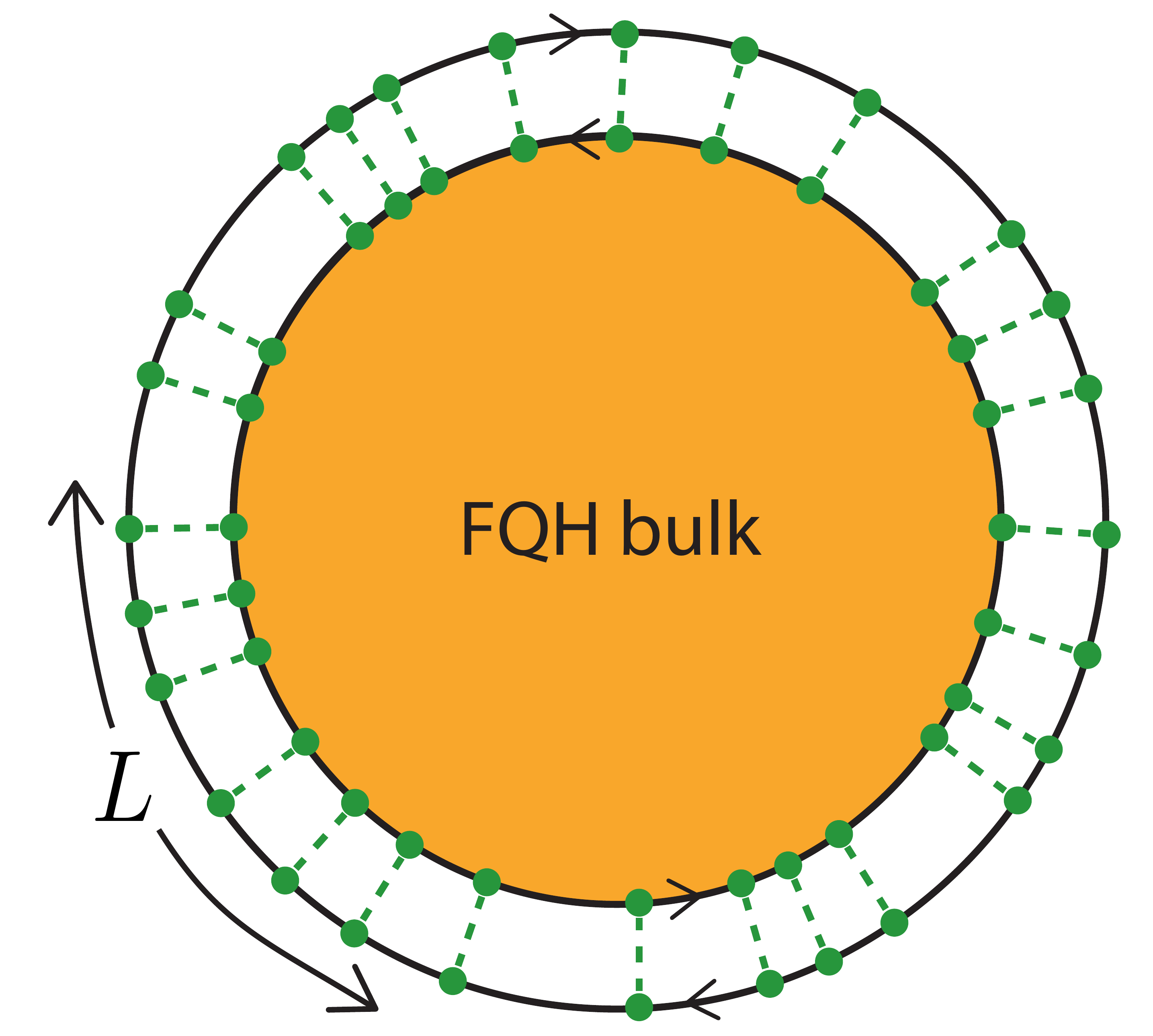}
\centering
\caption{Random impurity model: two counter-propagating chiral Luttinger liquids in a circular geometry of length $L$, together with $M$ randomly positioned impurity scatterers. }
\label{random_lattice}
\end{figure}

Our basic setup is as follows. We consider a circular edge of circumference $L$ with $M$ randomly positioned impurities. We denote the spacing between the impurities by $\ell_1, ..., \ell_M$, and the average spacing by $\bar{\ell} = L/M$ (Fig. \ref{random_lattice}). We show that this system supports two low energy phonon modes, one of which is neutral and one {of} which is charged, and neither of which is localized.

The first step in our analysis is to view the random system as an impurity lattice consisting {of} a single unit cell of length $L$. We can then carry over all of our results on impurity lattices by simply setting $\ell = L$, $m = M$, and $k =0$. In particular, if we make these substitutions in (\ref{eigeqgencell}), we obtain the eigenvalue equation
\begin{equation}
T_{sys}(E) \cdot \bpm A \\ B \epm = \bpm A \\ B \epm
\label{Tsyseig}
\end{equation}
where
\begin{equation}
T_{sys}(E) = T D(E \ell_M) T \cdots T D(E \ell_1)
\end{equation}
is the transfer matrix describing the entire system. As before, every solution $(E, A, B)$ to this eigenvalue equation defines a phonon creation/annihilation operator with energy $E$.

The next step is to solve the above eigenvalue equation in the limit $E \rightarrow 0$. We do this with the help of the following approximate expression for $T_{sys}$:
\begin{align}
T_{sys}(E) = \exp\Big[- \frac{iEL}{2} (W &+TWT)  \nonumber \\
&+ O(E^{\frac{4}{3}} L \bar{\ell}^{\frac{1}{3}}  \|W\|^{\frac{4}{3}}) \Big]
\label{Tsysapprox}
\end{align}
Here $\|W\|$ is defined as the magnitude of the largest eigenvalue of $W$ {(see Appendix \ref{randsysapprox} for a derivation).} 

To use (\ref{Tsysapprox}), we substitute it into (\ref{Tsyseig}) and neglect the error term. This approximation is justified at sufficiently low energies, i.e.,
\begin{align}
E \ll \frac{1}{L^{3/4} \bar{\ell}^{1/4} \|W\|}
\label{enregime}
\end{align}
The result of the substitution is:
\begin{equation}
\exp\left[-\frac{iEL}{2} (W +TWT) \right] \cdot \bpm A \\ B \epm = \bpm A \\ B \epm
\label{Tsyseig2}
\end{equation}
Next, we observe that the following commutator vanishes, as in Eq.~(\ref{vancomm}):
\begin{align*}
[T, W + TWT] = 0
\end{align*}
It follows that the matrix on the left hand side of (\ref{Tsyseig2}) has the same eigenvectors as $T$, namely $\bpm 1 \\ 1 \epm$, $\bpm k_1 \\ k_2 \epm$. Thus,
\begin{align}
\bpm A \\ B \epm = \bpm 1 \\ 1 \epm \text{ or } \bpm k_1 \\ k_2 \epm
\end{align}
Plugging these eigenvectors into (\ref{Tsyseig2}), we can extract the corresponding energies with straightforward linear algebra:
\begin{equation}
E_{\rho n} = v_\rho \cdot \frac{2 \pi n }{L}, \quad \text{ or } \quad E_{\sigma n} = v_\sigma \cdot \frac{2 \pi n }{L}
\label{Eform}
\end{equation}
where $v_\rho, v_\sigma$ are given by the formulas in (\ref{vrhosigma}) and $n = 1,2,...$, etc. 

We can now derive both of our claims about the low energy phonon modes --- namely (1) they are not localized and (2) one is charged and the other is neutral. To see that the low energy phonon modes are not localized, notice that the energy levels in (\ref{Eform}) are equally spaced with a spacing proportional to $1/L$: this level spacing indicates that the localization length $\xi$ is larger than the system size $L$ for any $E$ satisfying (\ref{enregime}). To see that the $\sigma$ mode is neutral, notice that the eigenvector $\bpm k_1 \\ k_2 \epm$ associated with the $\sigma$ mode is an eigenvector of $T$ with eigenvalue $-1$. As a result, the phonon creation/annihilation operators for this mode are of the form $a_{\sigma n} = \int dx (f_{\sigma n}(x) \partial_x \phi_1 + g_{\sigma n}(x) \partial_x \phi_2)$ where $f_{\sigma n}$ and $g_{\sigma n}$ alternate signs at each impurity. Like in section \ref{neutral_cond}, these alternating signs suppress the contribution of the $a_{\sigma n}$ operator to the coarse-grained density $\bar{\rho}$ since the even and odd spacings are equal on average. It follows that the $\sigma$ mode is neutral.

\subsubsection{{Finite $U$}} \label{subsubsec:randfinU}

We now consider the same setup as above, but with finite scattering strength $U$. Our main result is that the charge and neutral modes continue to persist at sufficiently large $U$.

Like the toy model, we study the effect of finite $U$ by adding appropriate correction terms to the $U \rightarrow \infty$ low energy theory. For the random impurity model, the latter theory can be read off from the phonon dispersion relations (\ref{Eform}): these expressions imply that the $U \rightarrow \infty$ low energy theory is a variant of $\overline{H}_{\text{eff}}$ (\ref{Hlowreal}) where the $\rho$ and $\sigma$ modes have velocities $v_\rho$ and $v_\sigma$ instead of $\bar{v}$.

Since the low energy theory is almost the same as for the toy model, most of our analysis of finite $U$ corrections can be repeated without change. As before, there are only two kinds of correction terms we need to worry about: $\partial_x \phi_\rho \partial_x \phi_\sigma$ and $e^{ \pm i \phi_\sigma} \partial_x \phi_\rho$. Also as before, both of these terms are generated by finite $U$ corrections, but with spatially dependent coefficients. The first term, $\partial_x \phi_\rho \partial_x \phi_\sigma$, appears in a combination of the form
\begin{equation}
\sum_j (-1)^j c_j \partial_x \phi_\rho  \partial_x \phi_\sigma(x_j)
\end{equation}
while $e^{\pm i \phi_\sigma} \partial_x \phi_\rho$ appears in a combination of the form
\begin{equation}
\sum_j d_j \cos(\phi_\sigma(x_j) - \beta_j) \partial_x \phi_\rho(x_j)
\end{equation}
{The only difference between these expressions and Eqs. (\ref{rs}) and (\ref{esr}) is that the coefficients $c_j, d_j$ are $j$-dependent. This inhomogeneity is expected since each impurity experiences a different local environment due to the random spacing.} 

{The rest of the argument is identical to the one for the toy model. As before,} the alternating signs in the first expression and the random\footnote{We assume that the $\alpha_j$ are random for this model.} $\beta_j$ phases in the second expression have the effect of suppressing these two perturbations, making them \emph{irrelevant} in the RG sense. Since these are the only perturbations that can hybridize the charge and neutral mode, we conclude that the charge and neutral mode structure persists at sufficiently large $U$, as claimed above.


\section{Conclusion}\label{sec:conclusion}

In this paper we have presented a microscopic derivation of the neutral mode in various FQH edges, including the $\nu = 2/3$ edge. Our derivation applies to a particular set of models which consist of two counter-propagating chiral Luttinger liquids together with a collection of discrete impurity scatterers. Our main result is an \emph{exact} solution of these models in the limit of infinitely strong impurity scattering. From this solution, we have explicitly shown that the low energy theory of these {systems} consists of decoupled charge and neutral modes. In addition we have shown that the charge and neutral modes survive at finite but sufficiently strong scattering {as long as this scattering has a random spatial dependence}.

It is interesting to circle back and compare our results with the original neutral mode analysis of Kane, Fisher, and Polchinski.\cite{kfandp} In that work, the authors studied a model similar to the random impurity model $H^{\text{gen}}$ (\ref{hamWithPCgen}) for the case $k_1 = 1$ and $k_2 = 3$, i.e. the $\nu = 2/3$ state. Instead of a discrete set of scatterers, Ref.~\onlinecite{kfandp} considered a continuum scattering term of the form $\int dx (\xi(x) e^{ik_1 \phi_1 + ik_2 \phi_2} + \text{H.c})$ where $\xi(x)$ is a Gaussian random variable with $\overline{\xi^*(x) \xi(x')} = U^2 \delta(x-x')$ for some $U$.\footnote{Here we have modified the notation of Ref.~\onlinecite{kfandp}, where $\overline{\xi^*(x) \xi(x')} = W \delta(x-x')$, so that it is consistent with this paper.} While this model is not identical to $H^{\text{gen}}$, it is similar enough that we can compare results on a qualitative level. From this comparison we can see that the two works consider different parameter regimes. Ref.~\onlinecite{kfandp} established the existence of a neutral mode for the case where $U$ is arbitrary but the velocity matrix $V$ has the special property that the edge theory has nearly decoupled charge and neutral modes in the \emph{absence} of electron scattering. In contrast, we derive the neutral mode for large $U$ but \emph{arbitrary} $V$. This difference in parameter regimes implies a conceptual difference between our two analyses: while Ref.~\onlinecite{kfandp} established the stability of the charge and neutral mode structure to small perturbations, we show that electron scattering can produce charge and neutral modes out of a system whose bare ($U = 0$) mode structure is completely different. In this sense, the results in this paper are complementary to those of Ref.~\onlinecite{kfandp}.

One of {the main achievements of this work} has been to show that our models capture a nontrivial effect of impurity scattering, {namely the emergent neutral mode}. But impurity scattering also has another important effect on FQH edges: it provides a mechanism for equilibrating the chemical potential of different edge modes. Such equilibration is a crucial property of multi-mode edges and in fact is necessary to explain their observed quantized Hall conductance.\cite{kfandp, buttiker1988absence,kane1995contacts} Thus, it is natural to ask whether our models capture this equilibration physics. The answer to this question depends on whether we consider finite or infinitely strong impurity scattering. In the case of finite scattering strength, we believe that our models do exhibit equilibration, as would be expected for any sufficiently generic system. On the other hand, in the case of infinite scattering strength, our models do not display equilibration since they are \emph{integrable} (in fact quadratic) in this limit. Thus, while the infinite scattering limit provides an exactly solvable model for the neutral mode, it does not provide a model for edge equilibration physics.

We envision several directions for future work. One direction would be to extend {our analysis} to systems with more than two edge modes, such as the Jain states with filling fraction $n/(2n \pm 1$) {or a $\nu = 2/3$ state with edge reconstruction.\cite{sabo2017edge}} Many of these states are predicted to have neutral modes based on the same kind of RG analysis as in the {original $\nu = 2/3$ proposal}.\cite{kane1995impurity,moore1998classification} 
{Similarly,} it would be interesting to apply our approach to systems with Majorana modes such as the anti-Pfaffian state.\cite{levin2007particle, lee2007particle} 

Another direction would be to study the $\nu = 4/5$ edge. This example is interesting { because, in our language, it } corresponds to the case $k_1 = 1$ and $k_2 = 5$, so in particular it has $k_2 - k_1 > 2$. As we mentioned earlier, when $k_2 - k_1$ is larger than $2$, the infinite scattering limit exhibits an extensive ground state degeneracy in addition to charge and neutral modes. {This degeneracy} poses basic challenges for determining whether the charge and neutral modes survive at finite scattering strength. {Thus, a new approach may be needed to understand this case.}


\subsection*{Acknowledgements}
We thank Sriram Ganeshan for helpful discussions. CH and ML are supported in part by the NSF under
grant No. DMR-1254741.


\begin{appendix}

\section{Degeneracy}\label{Degeneracy} 
In this paper, we have made heavy use of the fact that the low energy spectrum of our models is described by non-interacting phonons in the limit $U \rightarrow \infty$. This result is correct for $k_2 - k_1 = 2$, but, as we mentioned earlier, it is not quite right for $k_2 - k_1 > 2$ due to an additional degeneracy in the energy spectrum. In this appendix we derive an explicit formula for this degeneracy: for a circular edge with $2N$ impurities and $k_1\ne k_2$, we show that every phonon occupation state, including the ground state, has a degeneracy of
\beq
D=\left|\frac{k_2-k_1}{2}\right|^{N-1}
\label{degform}
\eeq
in the limit $U \rightarrow \infty$. Notice that $D$ grows exponentially with $N$ when $k_2 - k_1 > 2$, so the degeneracy is \emph{extensive} in this case. 

\subsection{General method for computing degeneracy}

We begin by reviewing a method for computing degeneracy which applies to any Hamiltonian of the form (\ref{genh}). This method was derived in Ref.~\onlinecite{quadham} and it goes as follows: the first step is to compute the commutator matrix
\beq
\mathcal{Z}_{ij}=\frac{1}{2\pi i}[C_i,C_j]
\eeq
The second step is to make a linear change of variables,\footnote{In Ref.~\onlinecite{quadham}, this change of variables includes an offset, i.e. $C_i' = \sum_j \mathcal{V}_{ij} C_j + \chi_i$, but we do not need to include $\chi_i$ here as it does not play a role in the degeneracy computation.} 
\beq
C_i'=\sum_{j} \mathcal{V}_{ij}C_j\non
\eeq
such that (i) $\mathcal{V}$ is an integer matrix with determinant $\pm 1$, and (ii) the matrix $\mathcal{Z}'=\frac{1}{2\pi i} [C_i', C_j']$ is in skew-normal form:
\beq
\mathcal{Z}'=
\bpm 0 & -\mathcal{D} &0 \\
\mathcal{D}  & 0 &0 \\
0& 0  &0 \epm 
\label{skewnorm}
\eeq
where
\beq
\mathcal{D}=
\bpm d_1 & 0 & \cdots & 0 \\
0 & d_2 & \cdots & 0 \\
\vdots & \vdots & \vdots & \vdots \epm
\label{dmatrix}
\eeq
and the $d_i$ are all nonzero. Such a change of variables always {exists}, although it is not necessarily unique. After making this change of variables, the degeneracy can be computed as
\beq
D= \left|\prod_{i=1}^N d_i \right|
\label{gendegform}
\eeq
The intuition behind this procedure is that the degeneracy arises because the arguments of the cosine terms, i.e. the $C_i$, do not commute with one another; hence to compute the degeneracy, we need to carefully analyze the commutation relations of the $C_i$. 
For more details, we refer the reader to Ref.~\onlinecite{quadham}.

\subsection{Application to impurity model}
We now compute the degeneracy of our system of $2N$ impurities arranged in a disk geometry. Before we start, we first need to take care of a technical issue. This issue is that the above method for computing degeneracy is designed for systems where all the degrees of freedom are continuous and real valued (e.g like $x$ and $p$) but our system has two degrees of freedom that take integer values, namely the total charge on each edge mode:
\begin{equation}
Q_i = \frac{1}{2\pi}\int dx \partial_x \phi_i, \quad i = 1,2
\end{equation}
Likewise, our system has two compact degrees of freedom that take values in $[0, 2\pi)$, namely $k_1 \phi_1$ and $k_2 \phi_2$. 

Fortunately, there is a trick for dealing with this discrepancy, which was introduced by Ref.~\onlinecite{quadham}. The trick is to treat all the degrees of freedom in our system as though they are real valued, and then enforce the quantization of $Q_1, Q_2$ and the compactness of $\phi_1, \phi_2$ at an energetic level by adding two more cosine terms to the Hamiltonian: 
\begin{equation*}
H \rightarrow H - U \cos(2\pi Q_1) - U \cos(2\pi Q_2)
\end{equation*}
In the limit $U \rightarrow \infty$, these cosine terms lock $Q_1, Q_2$ to integer values and also make the corresponding conjugate varables, $\phi_1, \phi_2$ compact.
 
With the help of this trick, it is straightforward to apply the above method to our system. All together, we have $2N+2$ cosine terms $\cos(C_j)$ with 
\begin{align*}
C_j &= k_1 \phi_1(x_j) + k_2 \phi_2(x_j) -\alpha_j, \quad j = 1,..., 2N \\
C_{2N+1} &= 2\pi Q_1, \quad C_{2N+2} = 2\pi Q_2
\end{align*}
To compute the corresponding commutator matrix $\mathcal{Z}_{ij}$, we need to fix a convention for the commutation relations of $\phi_1, \phi_2$. We use the following convention:
\begin{align*}
[\phi_1(x_i), \phi_1(x_j)] &= \frac{\pi i}{k_1} \text{sgn}(i-j) \nonumber \\
[\phi_2(x_i), \phi_2(y_j)] &= -\frac{\pi i}{k_2} \text{sgn}(i-j) \nonumber \\
\end{align*}

From the above commutation relations, we obtain
\begin{align}
\mathcal{Z}_{ij} = \bpm 0 & c & c & \cdots & c & -1 & 1 \\ 
-c & 0 & c & \cdots & c & -1 & 1 \\
-c & -c & 0 & \cdots & c & -1 & 1 \\
\vdots & \vdots & \vdots & \vdots & \vdots & \vdots & \vdots \\
-c & -c & -c & \cdots & 0 & - 1 & 1 \\
1 & 1 & 1 & \cdots & 1 & 0 & 0 \\
-1 & -1 & -1 & \cdots & -1 & 0 & 0 \epm
\end{align}
where $c = \frac{k_2 - k_1}{2}$. The next step is to find a change of variables $C_i' = \sum_j \mathcal{V}_{ij} C_j$ such that $\mathcal{Z}_{ij}' = \frac{1}{2\pi i}  [C_i', C_j']=\mathcal{V}Z\mathcal{V}^T$ is in skew-normal form (\ref{skewnorm}). One can check that following change of variables does the job:
\begin{align}
C_1'&=C_{2N+1}\non \\
C_m'&=C_{2m}-C_{2m-1}, \ \ m=2,...,N\non \\
C_{N+1}'&=C_2\non \\
C_{N+m}'&=(C_1-C_2)+\sum_{k=1}^{m-1}(C_{2k+1}-C_{2k}), \ m=2,..., N\non \\
C_{2N+1}'&=C_1-C_2+C_3-C_4...+C_{2N-1}-C_{2N}\non \\
& +\frac{1+k_1}{2}C_{2N+1}+\frac{1+k_2}{2}C_{2N+2}\non \\
C_{2N+2}'&= -C_1+C_2-C_3+C_4...-C_{2N-1}+C_{2N}\non \\ 
&+\frac{1-k_1}{2}C_{2N+1}+\frac{1-k_2}{2}C_{2N+2}
\label{change-of-variables}
\end{align}
The corresponding $\mathcal D$ matrix in (\ref{skewnorm}) has dimension $N \times N$ with diagonal entries
\beq
d_1=-1, \quad d_i=\frac{k_2-k_1}{2}; \quad  i=2,\cdots,N\non
\eeq
Substituting these values into the general formula for the degeneracy (\ref{gendegform}) gives $D =  \left|\frac{k_2-k_1}{2}\right|^{N-1}$. This completes our derivation of (\ref{degform}).


\section{Regularizing the impurity scattering terms} \label{regularized-delta}

In this appendix, we derive Eq. (\ref{constraint2b}) from the constraint $[a, C_j] = 0$ by appropriately regularizing the impurity scattering terms. Our derivation closely follows a similar appendix in Ref.~\onlinecite{quadham}. 

To see why we need to regularize at all, suppose we directly substitute the definition of $a$ (\ref{aop}) into $[a,C_j] = 0$ and evaluate the commutator. The result is:
\begin{equation}
f(j \ell) = g(j \ell)
\end{equation}
It is hard to make sense of this equation since the expressions for $f$ and $g$ (\ref{fgdef}) are discontinuous at $x = j \ell$ and hence $f(j \ell)$ and $g(j \ell)$ are not well-defined. What we will show below is that regularizing changes the above equation to the more sensible relation
\begin{equation}
\frac{f(j \ell^-)+f(j \ell^+)}{2} = \frac{g(j \ell^-)+g(j \ell^+)}{2}
\end{equation}
  
Our regularization scheme is as follows: for each impurity scattering term $\cos(C_j)$, we replace $C_j = k_1 \phi_1(j \ell) + k_2 \phi_2(j \ell) -\alpha_j$ with
\beq
C_j= \int_{-\infty}^{\infty}dx\ \tilde{\delta}(x-j \ell)[k_1\phi_1(x)+k_2\phi_2(x)] -\alpha_j 
\eeq
where $\tilde{\delta}(x)$ is an approximation to a delta function, i.e. a narrowly peaked function with $\int\tilde{\delta}(x) dx =1$. One can think of this replacement as effectively introducing a short distance cutoff into our model. 

Once we make this substitution, we repeat the calculation in Eqs. (\ref{fgdiff} - \ref{fgdef}) and solve for the functions $f$ and $g$. We obtain  
{
\begin{align}
&f(x)=\sum_{j} A^{(j)}e^{-i\frac{E}{v}(x-j \ell)}\Big[\tilde{\Theta}_1(x-j \ell)-\tilde{\Theta}_1(x-(j+1)\ell)\Big]\non\\
&g(x)=\sum_{j} B^{(j)}e^{i\frac{E}{v}(x-j \ell)}\Big[\tilde{\Theta}_2(x-j \ell)-\tilde{\Theta}_2(x-(j+1)\ell)\Big]\non
\end{align}
}
where $\tilde{\Theta}_1$ and $\tilde{\Theta}_2$ are regularized versions of the Heaviside step function:
\begin{align}
\tilde{\Theta}_1(x) &= \int_{-\infty}^x e^{i\frac{E}{v}y} \tilde{\delta}(y) dy \nonumber \\ 
\tilde{\Theta}_2(x) &= \int_{-\infty}^x e^{-i\frac{E}{v}y} \tilde{\delta}(y) dy 
\end{align}
Next we note that the constraint $[a,C_{j}]=0$ gives \beq\label{constraintAC}
\int_{-\infty}^{\infty}dx\ [f(x)-g(x)]\tilde{\delta}(x-j \ell)=0
\eeq
To complete the calculation, we need to substitute the above expressions for $f$ and $g$ into (\ref{constraintAC}) and evaluate the resulting integral. We do this with the help of the following identity: {
\begin{align}
&\lim_{ \frac{E}{v}b\rightarrow 0}\int_{-\infty}^{\infty}dx\ \tilde{\Theta}_s(x-j \ell) \tilde{\delta}(x-j'\ell)  e^{\pm i\frac{E}{v}x} \nonumber \\
&= \begin{cases}
0 & \ j > j' \\
e^{\pm i \frac{E}{v} j' \ell} & \ j < j' \\
\frac{1}{2} e^{\pm i\frac{E}{v}j \ell} & \ j = j'
\end{cases}
\end{align}
Here $b$ is the characteristic width of the $\tilde{\delta}(x)$ function and $s$ runs over the two values $s = 1,2$. The justification for this identity for $j > j'$ and $j < j'$ is obvious; as for $j =j'$, we can prove it for $s = 1$} by noting that
\begin{align}
&\lim_{ \frac{E}{v}b\rightarrow 0}\int_{-\infty}^{\infty}dx\ \tilde{\Theta}_1(x-j \ell) \tilde{\delta}(x-j \ell)  e^{\pm i\frac{E}{v}x}\non \\
&=\lim_{ \frac{E}{v}b\rightarrow 0}\int_{-\infty}^{\infty}dx\int_{-\infty}^{x}dy\  \tilde{\delta}(y-j \ell) \tilde{\delta}(x-j \ell) e^{i \frac{E}{v}(y-j \ell) \pm i\frac{E}{v}x}\non \\
&=e^{\pm i\frac{E}{v}j \ell}\lim_{ \frac{E}{v}b\rightarrow 0}\int_{-\infty}^{\infty}dx\int_{-\infty}^{x}dy\ \tilde{\delta}(y-j \ell) \tilde{\delta}(x-j \ell) \non \\
&= \frac{1}{2} e^{\pm i\frac{E}{v}j \ell}\lim_{ \frac{E}{v}b\rightarrow 0}\int_{-\infty}^{\infty}dx\int_{-\infty}^{\infty}dy\ \tilde{\delta}(y-j \ell) \tilde{\delta}(x-j \ell) \non \\
&=\frac{1}{2}e^{\pm i\frac{E}{v}j \ell}
\end{align}
The proof for {$s = 2$} is similar. 

Applying the above identity to (\ref{constraintAC}) and simplifying, we arrive at the condition
\beq
\frac{A^{(j)}+A^{(j-1)}e^{-i\frac{E \ell}{v}}}{2}=\frac{B^{(j)}+B^{(j-1)}e^{i\frac{E \ell}{v}}}{2}
\label{constraint2b-appendix}
\eeq
This is exactly Eq. \ref{constraint2b}, which we wished to derive.

\section{Deriving the approximation (\ref{Tsysapprox})} \label{randsysapprox}
In this appendix we derive Eq. (\ref{Tsysapprox}), which gives an approximate expression for the transfer matrix $T_{sys}$ 
for a system of $M$ impurities randomly arranged on a circular edge of circumference $L$. As in the main text, we denote the spacing between the impurities by $\ell_1,...,\ell_M$ so that
\begin{equation*}
T_{sys}(E) = T D(E \ell_M) T \cdots T D(E \ell_1)
\end{equation*}
with $D(x) = e^{-i W x}$ and $W = K V^{-1}$.

For simplicity, we will assume that the number of impurities $M$ is a power of $2$. This allows us to factor $M$ as $M = r\cdot (M/r)$ where $r$ is a smaller power of $2$. We can then write $T_{sys}$ as a product of $(M/r)$ terms, each of which involves $r$ impurities. That is:
\begin{equation}
T_{sys}(E) = T_{M/r}(E) \cdots T_{2}(E) \cdot T_{1}(E)
\label{Tsysdec}
\end{equation}
where
\begin{align*}
T_1(E) &= T D(E \ell_{r}) T \cdots T D( E \ell_{1}) \\
T_2(E) &= T D(E \ell_{2r}) T \cdots T D( E \ell_{r+1}) \\
\vdots
\end{align*}
and so on. For the moment, we will leave the value of $r$ unspecified; later we will choose $r$ so as to obtain the best bound on the error in our approximations. 

Next, we expand each $T_j(E)$ to linear order in $E$. Using the fact that $T^2 = \mathbbm{1}$, this gives
\begin{equation*}
T_j(E) \approx \mathbbm{1} - i E ( W \ell_{even,j} + TWT \ell_{odd,j})
\end{equation*}
where
\begin{align*}
\ell_{odd,j} &= \ell_{jr-r+1} + \ell_{jr-r+3} + \dots + \ell_{jr-1} \\
\ell_{even,j} &= \ell_{jr-r+2} + \ell_{jr-r+4} + \dots + \ell_{jr}
\end{align*}
For a typical impurity distribution, the even and odd spacings are approximately equal: 
\begin{align*}
\ell_{even,j} \approx \ell_{odd,j} \approx \frac{r \bar{\ell}}{2}
\end{align*}
Hence, the above expression for $T_j(E)$ can be simplified to
\begin{equation}
T_j(E) \approx \mathbbm{1} - \frac{r i E \bar{\ell}}{2} [W + TWT] 
\end{equation}
Let us try to bound the total error in the above approximation. There are two errors we need to think about: the systematic error coming from expanding $T_j(E)$ to linear order in $E$ and the statistical error coming from replacing $\ell_{even,j}$ and $\ell_{odd,j}$ by their typical value, $r \bar{\ell}/2$. The systematic error can be estimated by the quadratic term in the expansion of $T_j(E)$, which is of order $O(r^2 E^2 \bar{\ell}^2 \|W \|^2 )$ where $\|W\|$ is the magnitude of the largest eigenvalue of $W$. As for the statistical error, we expect this to be proportional to the typical size of the fluctuations in $\ell_{even,j}$ and $\ell_{odd,j}$, which are both of order $\sqrt{r} \bar{\ell}$, so we obtain the estimate $O(\sqrt{r} E\bar{\ell} \|W\|)$. To get an optimal bound on the total error, we choose $r$ so that these two errors have the same size, i.e. 
\begin{equation*}
r \sim  (E  \bar{\ell} \|W\|)^{-\frac{2}{3}}
\end{equation*}
For this choice of $r$, both errors are of order $O( E^{\frac{2}{3}}  \bar{\ell}^{\frac{2}{3}} \|W\|^{\frac{2}{3}})$, so that
\begin{equation}
T_j(E) = \mathbbm{1} - \frac{r i E \bar{\ell}}{2} (W + TWT) + O(E^{\frac{2}{3}} \bar{\ell}^{\frac{2}{3}} \|W\|^{\frac{2}{3}}  )
\label{Tjapprox}
\end{equation}

Substituting the above expression (\ref{Tjapprox}) into (\ref{Tsysdec}), we derive 
{
\begin{align}
&T_{sys}(E) = \left[\mathbbm{1} - \frac{r i E \bar{\ell}}{2} (W + TWT) +O(E^{\frac{2}{3}} \bar{\ell}^{\frac{2}{3}} \|W\|^{\frac{2}{3}} ) \right]^{\frac{M}{r}} \nonumber \\
&= \exp \left[- \frac{i M E \bar{\ell}}{2} (W +TWT) + O(M E^{\frac{4}{3}} \bar{\ell}^{\frac{4}{3}}  \|W\|^{\frac{4}{3}}) \right] \nonumber \\
&= \exp \left[- \frac{iEL}{2} (W +TWT) + O(E^{\frac{4}{3}} L \bar{\ell}^{\frac{1}{3}}  \|W\|^{\frac{4}{3}}) \right]
\end{align}
}
which is exactly Eq. (\ref{Tsysapprox}). We note that the error term in the above approximation is an \emph{upper bound} and is likely larger than the true error. 


\end{appendix}
\bibliography{edgemodes}


\end{document}